\documentclass[twocolumn, tighten, times, twocolappendix]{aastex7}

\begin{document}

\title{A Spectroscopic Search for Dormant Black Holes in Low-Metallicity Binaries}

\author[orcid=0000-0002-1386-0603,gname=Pranav,sname=Nagarajan]{Pranav Nagarajan}
\affiliation{Department of Astronomy, California Institute of Technology, 1200 E. California Blvd., Pasadena, CA 91125, USA}
\email[show]{pnagaraj@caltech.edu}  

\author[orcid=0000-0002-6871-1752,gname=Kareem,sname=El-Badry]{Kareem El-Badry}
\affiliation{Department of Astronomy, California Institute of Technology, 1200 E. California Blvd., Pasadena, CA 91125, USA}
\email{kelbadry@caltech.edu}  

\author[orcid=0000-0001-6533-6179,gname=Henrique,sname=Reggiani]{Henrique Reggiani}
\affiliation{Gemini Observatory/NSF’s NOIRLab, Casilla 603, La Serena, Chile}
\email{henrique.reggiani@noirlab.edu}

\author[orcid=0000-0002-6406-1924,gname=Casey,sname=Lam]{Casey Y. Lam}
\affiliation{Observatories of the Carnegie Institution for Science, 813 Santa Barbara St., Pasadena, CA 91101, USA}
\email{clam@carnegiescience.edu}

\author[orcid=0000-0002-4733-4994,gname=Joshua,sname=Simon]{Joshua D. Simon}
\affiliation{Observatories of the Carnegie Institution for Science, 813 Santa Barbara St., Pasadena, CA 91101, USA}
\email{jsimon@carnegiescience.edu}

\author[orcid=0000-0001-9590-3170,gname=Johanna,sname=M\"uller-Horn]{Johanna M\"uller-Horn}
\affiliation{Max Planck Institute for Astronomy, K\"onigstuhl 17, D-69117, Heidelberg, Germany}
\email{mueller-horn@mpia.de}

\author[orcid=0000-0001-8898-9463,gname=Rhys,sname=Seeburger]{Rhys Seeburger}
\affiliation{Max Planck Institute for Astronomy, K\"onigstuhl 17, D-69117, Heidelberg, Germany}
\email{seeburger@mpia.de}

\author[orcid=0000-0003-4996-9069,gname=Hans-Walter,sname=Rix]{Hans-Walter Rix}
\affiliation{Max Planck Institute for Astronomy, K\"onigstuhl 17, D-69117, Heidelberg, Germany}
\email{rix@mpia.de}

\author[orcid=0000-0002-0531-1073,gname=Howard,sname=Isaacson]{Howard Isaacson}
\affiliation{Department of Astronomy, University of California, Berkeley, Berkeley, CA, 94720, USA}
\email{hisaacson@berkeley.edu}

\author[orcid=0000-0001-9611-0009,gname=Jessica,sname=Lu]{Jessica R. Lu}
\affiliation{Department of Astronomy, University of California, Berkeley, Berkeley, CA, 94720, USA}
\email{jlu.astro@berkeley.edu}

\author[orcid=0000-0002-0572-8012,gname=Vedant,sname=Chandra]{Vedant Chandra}
\affiliation{Center for Astrophysics $|$ Harvard \& Smithsonian, 60 Garden Street, Cambridge, MA 02138, USA}
\email{vedant.chandra@cfa.harvard.edu}

\author[orcid=0000-0001-8006-6365,gname=Rene,sname=Andrae]{Rene Andrae}
\affiliation{Max Planck Institute for Astronomy, K\"onigstuhl 17, D-69117, Heidelberg, Germany}
\email{andrae@mpia.de}

\begin{abstract}

The discovery of the massive black hole (BH) system Gaia BH3 in pre-release \textit{Gaia} DR4 data suggests that wide BH binaries with luminous companions may be significantly overrepresented at low metallicities. Motivated by this finding, we have initiated a spectroscopic survey of low-metallicity stars exhibiting elevated \texttt{RUWE} values in {\it Gaia} DR3, using the FEROS and APF spectrographs. We identify promising BH binary candidates as objects with instantaneously measured radial velocities (RVs) that are very different from their mean RVs reported in \textit{Gaia} DR3. Thus far, we have observed over 500 targets, including a nearly complete sample of stars with $\rm[Fe/H] < -1.5$, \texttt{RUWE} $> 2$, and $G < 15$. Our search has yielded one promising target exhibiting slow acceleration and an RV more than 98 km s$^{-1}$ different from its DR3 mean RV, as well as dozens of other candidates with smaller RV discrepancies. We quantify the sensitivity of our search using simulations, demonstrating that it recovers at least half of the BH companions within our selection criteria. We make all the spectra and RVs from our survey publicly available and encourage further follow-up.

\end{abstract}

\keywords{\uat{Stellar astronomy}{1583}, \uat{Black holes}{162}}


\section{Introduction}
\label{sec:intro}

\begin{figure*}
    \centering
    \includegraphics[width=\textwidth]{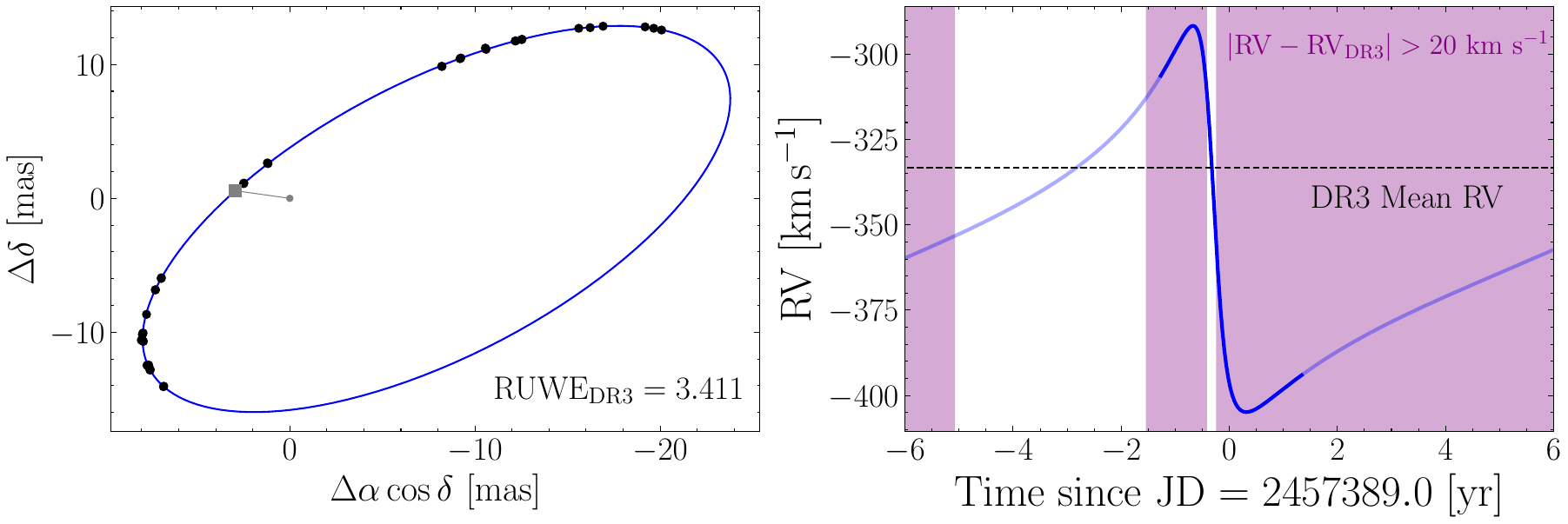}
    \caption{Simulated \textit{Gaia} DR3 epoch astrometry and radial velocity curve for Gaia BH3 based on best-fit combined parameters from \citet{gaia_collaboration_discovery_2024}. In the left panel, the gray line connects the barycenter to periastron. In the right panel, the temporal range of the simulated DR3 epoch astrometry is plotted with higher opacity. The epoch RVs of Gaia BH3 are different from its reported mean RV by at least 20 km s$^{-1}$ over $\sim70\%$ of its orbit (shaded purple region), suggesting that a search strategy based on this comparison may be useful in finding similar targets.}
    \label{fig:bh3_rv_astro}
\end{figure*}

The Milky Way (MW) is thought to contain $\sim10^8$ stellar-mass black holes \citep[BHs; e.g.,][]{olejak_synthetic_2020}. Only a tiny fraction of this population has been discovered, as BH searches are severely affected by observational incompleteness. For five decades, the only known Galactic BHs were in close binaries that shine brightly in X-rays due to accretion \citep[e.g.,][]{corral-santana_blackcat_2016}. However, these X-ray binaries are a rare outcome of binary evolution, and are expected to be vastly outnumbered by both wider binaries hosting non-accreting (``dormant'') BHs \citep[e.g.,][]{portegies_zwart_formation_1997, chawla_gaia_2022} and isolated BHs \citep[e.g.,][]{lam_isolated_2022, sahu_isolated_2022}. 

The {\it Gaia} mission \citep{gaia_collaboration_gaia_2016} has opened a new window on the population of rare binaries in the Milky Way, with the mission’s 3rd data release (``DR3", \citealt{gaia_collaboration_gaia_2023}) increasing the number of well-characterized binary orbits by more than an order of magnitude \citep{gaia_collaboration_gaia_2023-1}. \textit{Gaia} produces catalogs of photometrically, spectroscopically, and astrometrically variable binaries \citep{gaia_collaboration_gaia_2023-1, mowlavi_gaia_2023, gomel_gaia_2023, eyer_gaia_2023, halbwachs_gaia_2023}, and several recent studies have searched for dormant BHs in these catalogs \citep[e.g.,][]{el-badry_what_2022, shahaf_triage_2023, nagarajan_spectroscopic_2023}. In contrast to most other methods, \textit{Gaia} astrometry is most sensitive to binaries with relatively wide orbits \citep[e.g.,][]{halbwachs_gaia_2023, el-badry_gaias_2024}. Notably, DR3 astrometry has led to the discovery and characterization of two nearby dormant BHs (``Gaia BHs'',  \citealt{el-badry_sun-like_2023, chakrabarti_noninteracting_2023, nagarajan_espresso_2024, tanikawa_2023, el-badry_red_2023}) and a population of dormant neutron stars (NSs) (``Gaia NSs'', \citealt{el-badry_19_2024, el-badry_population_2024}) in au-scale orbits. Population synthesis studies calibrated to the DR3 detections predict discovery of dozens more Gaia BHs in future data releases \citep[e.g.,][]{nagarajan_realistic_2025}.

Recently, \citet{gaia_collaboration_discovery_2024} announced the discovery of Gaia BH3, a $33\,M_{\odot}$ BH orbited by a $0.8\,M_{\odot}$ red giant in an $11.6$ yr orbit, in pre-release astrometry from the mission's 4th data release (``DR4''). Gaia BH3 hosts the most massive known stellar-mass BH in the Milky Way, and the lowest-metallicity (${\rm [Fe/H]} = -2.56$, see also \citealt{balbinot_2024}) luminous companion to a BH, providing empirical evidence that low-metallicity massive stars leave behind massive BHs. Because Gaia BH3 is in the typical mass range of the merging BHs detected via gravitational waves (GWs) at extragalactic distances \citep[e.g.,][]{abbott_ligo_2023}, the system also provides tantalizing evidence that low-metallicity massive stars are the progenitors of GW events.

Since Gaia BH3 is relatively bright ($G = 11.2$), there may be many similar, but fainter, systems waiting to be discovered. Indeed, for a Kroupa initial mass function (IMF) \citep{kroupa_2001}, there are $\sim 10^2$ times more low-mass metal-poor main sequence stars in our solar neighborhood than there are red giants similar to the companion star in Gaia BH3. Hence, even within 590 pc (the distance to Gaia BH3), there likely exist other massive BHs that are orbited by metal-poor stars still on the main sequence. At the distance and extinction of Gaia BH3, such companions would have $G < 15$ for $M \gtrsim 0.67\,M_{\odot}$, readily accessible to \textit{Gaia} astrometry and spectroscopic follow-up. 

Even though we do not know the exact selection function that yielded the three known Gaia BHs, luminous companions to Gaia BHs (and Gaia NSs, see \citealt{el-badry_population_2024}) appear to be significantly overrepresented at low metallicity \citep{el-badry_gaiabh3_2024}. Indeed, based on the \text{Gaia} XP metallicity catalog of \citet{andrae_2023}, only $1$ in $10^4$ giants in the solar neighborhood is as metal-poor as Gaia BH3, and Gaia BH3 is among the $\approx 25$ nearest giants of such low metallicity. Why wide BH and NS companions might be overrepresented at low metallicity is an open question. One possibility is that low-metallicity massive stars remain more compact in their post-main-sequence evolution than solar-metallicity stars of the same initial mass \citep[e.g.,][]{maeder_1987, marchant_2016, klencki_2020}. While solar-metallicity BH progenitors expand to become red supergiants and engulf their stellar companions, very low-metallicity massive stars may not expand as much, allowing a larger fraction of binaries to survive and producing more BH + low-mass star binaries \citep[e.g.,][]{iorio_boring_2024}.

Previous studies have exhaustively searched the catalog of DR3 orbital solutions for dormant BHs and NSs in wide orbits \citep{andrews_sample_2022, shahaf_triage_2023, el-badry_sun-like_2023, el-badry_red_2023, el-badry_19_2024, el-badry_population_2024}. In this work, we perform spectroscopic follow-up on a sample of low-metallicity stars with acceleration solutions or high astrometric excess noise (Renormalized Unit Weight Error, or ``\texttt{RUWE}'') values in \textit{Gaia} DR3, indicating binarity. This investigation cannot be carried out with \textit{Gaia} data alone. \textit{Gaia} DR3 includes mean radial velocities (RVs) for stars with $G_{\text{RVS}} \lesssim 14$, but these represent an average RV measured from spectra coadded over several years \citep{katz_gaia_2023}. Epoch-level RVs are only measured for stars with $G_{\text{RVS}} \lesssim 12$, and are not included in the DR3 catalog \citep{katz_gaia_2023}. External, time-resolved RVs are thus crucial for constraining the possible nature of the companions.

In Section \ref{sec:motivation}, we describe and motivate our search method. In Section \ref{sec:sample}, we detail our sample selection process. In Section \ref{sec:data}, we describe the spectra obtained from our follow-up campaign. In Section \ref{sec:analysis}, we analyze the collected RVs and identify promising compact object candidates for further follow-up. We discuss the completeness and expected yields of our search in Section \ref{sec:discussion}. Finally, we summarize our results in Section \ref{sec:conclusion}.

\section{Search Strategy}
\label{sec:motivation}

\begin{figure*}
    \centering
    \includegraphics[width=\textwidth]{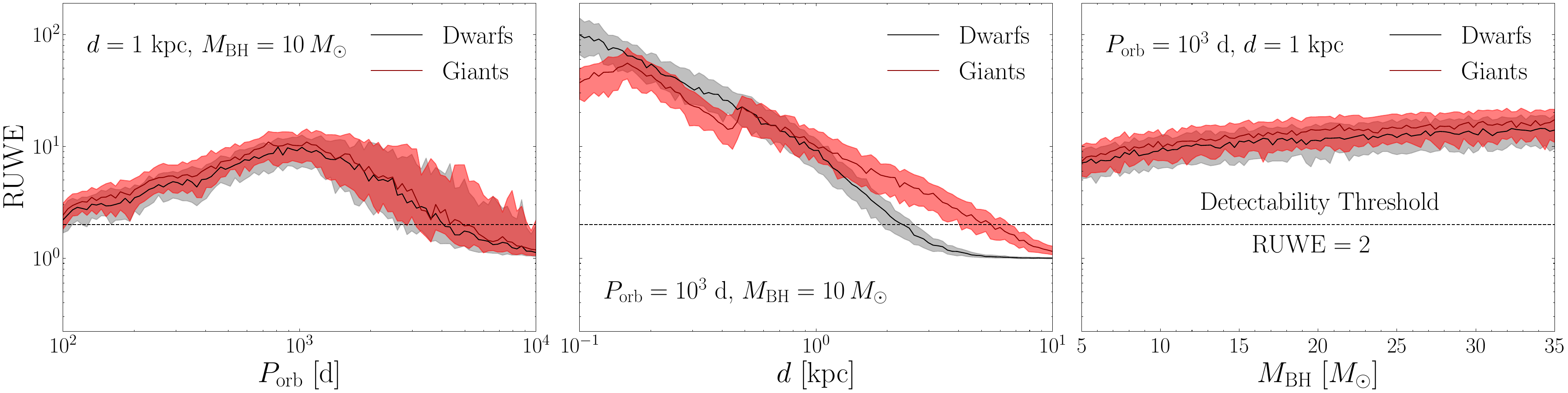}
    \caption{Dependence of predicted \texttt{RUWE} in DR3 on orbital period, distance, and BH mass for low-metallicity dwarf or giant companions to BHs, with the shaded regions representing a 68\% confidence interval. We randomly sample sky positions from our catalog of candidates, orbital eccentricities from a thermal distribution, and isotropic binary orientations. At orbital periods of $10^2$--$10^3$ d, \texttt{RUWE} increases as the projected semi-major axis of the photocenter increases, peaking at $\sim10^3$ days. At orbital periods $> 10^3$ d, the limited orbital coverage during \textit{Gaia} DR3's $\sim1000$ d observing baseline causes \texttt{RUWE} to decrease. \texttt{RUWE} also decreases as the binary gets farther away, since the projected semi-major axis of the photocenter decreases. \texttt{RUWE} is only weakly sensitive to BH mass. Due to their brighter absolute magnitudes, giants tend to have slightly larger \texttt{RUWE} values than dwarfs, and are detectable out to farther distances. With a selection cut of \texttt{RUWE} $> 2.0$, we are sensitive to orbital periods of $100$--$5000$ d and distances out to $\sim2$ kpc.}
    \label{fig:ruwe_trend_fig}
\end{figure*}

We design our search to be sensitive to Gaia BH3-like binaries. We use the \texttt{gaiamock} code \citep{el-badry_generative_2024}, which forward-models \textit{Gaia} epoch astrometry and model fitting, to simulate epoch astrometry and a radial velocity (RV) curve for Gaia BH3 as it would have been observed in \textit{Gaia} DR3. We show the epoch astrometry and RV curve in Figure~\ref{fig:bh3_rv_astro}, with times provided relative to the DR3 reference epoch of Julian Date (JD) $= 2457389.0$. The epoch astrometry was insufficiently constraining for Gaia BH3 to receive an orbital solution in DR3. However, the source had \texttt{RUWE} $= 3.4$, indicative of a poor single-star astrometric fit \citep{lindegren_ruwe_2018}. In addition, the star's RV is predicted to be different from the reported mean RV in DR3 by $\geq 20$ km s$^{-1}$ over $\sim70\%$ of its orbit. This suggests that a search strategy based on measuring epoch RVs for low-metallicity stars with high \texttt{RUWE} values in DR3, and comparing those RVs to the corresponding mean RVs reported in DR3, could lead to the discovery of BH or NS binaries that did not receive orbital solutions in DR3. Indeed, RV follow-up complements and validates \texttt{RUWE}-based filtering, with even a few RVs being able to break the degeneracy between low companion mass, long-period systems and high companion mass, short-period systems that both have elevated \texttt{RUWE}. The interpretation of any observed RV discrepancy between a binary's epoch RVs and its mean DR3 RV depends on its true orbital period (see Section \ref{sec:sensitivity}).

To explore what types of BH binaries can be detected via elevated \texttt{RUWE}, we investigate the variation of \texttt{RUWE} with orbital period, distance, and BH mass. We randomly sample sky positions from our catalog of candidate low-metallicity binaries (see Section \ref{sec:sample}), orbital eccentricities from a thermal distribution, and isotropic binary orientations. We adopt default values of $P_{\text{orb}} = 10^3$~d, $d = 1$ kpc, and $M_{\text{BH}} = 10\,M_{\odot}$, typical of binaries that our search strategy is sensitive to, and predict the \texttt{RUWE} in DR3 using \texttt{gaiamock}. We vary one quantity at a time in each panel in Figure~\ref{fig:ruwe_trend_fig}. We plot curves separately for dwarfs and giants, assuming absolute magnitudes of $M_G = 4$ for dwarfs and $M_G = 0$ for giants. Initially, \texttt{RUWE} increases as the orbital period and projected semi-major axis of the photocenter increase, peaking at $\sim10^3$ days. At $> 10^3$ d, however, DR3's limited observing baseline causes \texttt{RUWE} to decrease with orbital period instead. \texttt{RUWE} also decreases as the distance to a binary increases, since this causes the projected semi-major axis of the photocenter to decrease. \texttt{RUWE} is only weakly sensitive to BH mass. 

In general, giants lead to slightly larger \texttt{RUWE} values than dwarfs, though there is a somewhat non-intuitive dependence of \texttt{RUWE} on distance in Figure~\ref{fig:ruwe_trend_fig}. This behavior is determined by how the epoch astrometry uncertainties vary with apparent $G$-band magnitude. The uncertainties are smallest at $G = 8$--$14$, and rise at brighter or fainter magnitudes outside this range \citep[][]{holl_astrometric_2023}. As sources become too bright, uncertainties rise and \texttt{RUWE} falls, causing giants to have smaller \texttt{RUWE} values than dwarfs at close distances. Companions to giants are more detectable than companions to dwarfs only at large distances ($d \gtrsim 1.5$ kpc), because the uncertainties for dwarfs increase rapidly at those distances.

Our main sample includes a selection cut of \texttt{RUWE} $>2.0$ (see Section \ref{sec:sample} for details). This implies that we are sensitive to orbital periods of $100$--$5000$ d and distances out to $\sim2$ kpc. That being said, binaries that have elevated \texttt{RUWE} can show no detectable RV discrepancy if they have high eccentricities (depending on DR3 phase coverage) or face-on orbits. We quantify this in Section~\ref{sec:discussion}.

\section{Sample Selection} 
\label{sec:sample}

We select candidate metal-poor binaries using two sets of simple cuts on \texttt{RUWE} and the metallicity. We adopt the data-driven metallicities derived by \citet{andrae_2023}, who use the XGBoost algorithm \citep{chen2016xgboost} to infer stellar parameters from low-resolution \textit{Gaia} XP spectra. These metallicities are reliable because \citet{andrae_2023} train their model on externally derived, high-fidelity stellar parameters from APOGEE \citep{apogee_dr17_2022} and augment their training data with a set of very-metal-poor stars that have $\rm [M/H]$ values similar to those in our sample. \citet{andrae_2023} also use cross-validation to demonstrate the reliability of their model predictions.

\subsection{$\rm [M/H] < -1.5$ sample}

We selected all sources from the \citet{andrae_2023} catalog with $\rm [M/H] < -1.5$, $\texttt{RUWE} > 2.0$, and the following quality cuts: 

\begin{itemize}
    \item Apparent magnitude $G< 15$, to allow for RV measurement with small telescopes. 
    \item parallax $\varpi > 0.5$ mas, corresponding to a distance limit of $\lesssim 2$ kpc. 
    \item $\texttt{ipd\_frac\_multi\_peak} < 3$, to exclude marginally resolved wide binaries with elevated \texttt{RUWE} due to blending rather than orbital motion \citep{lindegren_binarity_2022}. 
    \item Color $0.3 < G_{\rm BP}-G_{\rm RP} < 2.5$, to exclude hot and cool stars for which the metallicities from \citet{andrae_2023} are unlikely to be reliable. 
    \item $v_{\perp} > 70\,{\rm km\,s^{-1}}$, where $v_{\perp} = 4.74\,{\rm km\,s^{-1}}\times (\mu / \text{mas yr$^{-1}$})/(\varpi/\text{mas})$ is the source's tangential velocity, and $\mu$ is the source's total proper motion. This excludes sources with disk-like kinematics, which are more likely to be metal-rich contaminants. 
    \item A measured radial velocity in {\it Gaia} DR3, so that an RV change can be measured. 
    \item $\texttt{rv\_amplitude\_robust} > 15\,{\rm km\,s^{-1}}$, excluding sources with small RV shifts measured during the DR3 observing baseline. This cut applies only to the minority of sources with $G_{\rm RVS} < 12$, for which epoch RVs were measured in DR3. 
\end{itemize}

These cuts yielded a total of 595 sources. Of these, 139 already have an orbital solution reported in the \texttt{gaiadr3.nss\_two\_body\_orbit} catalog. Most of these are astrometric \texttt{Orbital} solutions and have been investigated elsewhere \citep[e.g.,][]{tanikawa_2023, el-badry_red_2023}. None imply massive dark companions. 

We measured at least one RV for 425 of the remaining 456 candidates. We have thus observed $\approx 93$\% of the remaining 456 objects in the $[\rm M/H]<-1.5$ sample, or $\approx95$\% of 595 when we include the objects with orbital solutions. Of the 456 candidates lacking orbital solutions, 108 have acceleration or variable acceleration solutions reported in the \texttt{gaiadr3.nss\_acceleration\_astro} catalog. 102 of these sources are in our observed sample. The acceleration solutions suggest that these are long-period ($\gtrsim 500$ d; \citealt{el-badry_generative_2024}) binaries, but are insufficient to constrain the companion masses on their own.

\subsection{$[\rm M/H]< -1$, large $a_0$ sample}

Once most of the $[\rm M/H] < -1.5$ sample had been observed, we selected additional metal-poor candidates with a selection designed to prioritize large photocenter orbits. We began by selecting sources with  $[\rm M/H] < -1.0$ according to \citet{andrae_2023} and  $\texttt{RUWE} > 1.4$, and applied the same cuts on $G$, $\varpi$, $v_{\perp}$, \texttt{ipd\_frac\_multi\_peak},  DR3 RV, color, and \texttt{rv\_amplitude\_robust} that are used in the $[\rm M/H]< -1.5$ sample. We removed any overlapping sources that were already present in the $[\rm M/H] < -1.5$ sample.

Motivated by \citet{el-badry_generative_2024} (their Figure 8), we predict the approximate photocenter orbit semi-major axis of each binary from \texttt{RUWE} as
\begin{equation}
    \label{eq:a0}
    \hat{a}_0 = 2.4 \sigma_\eta(G) \sqrt{\texttt{RUWE}^2 -1},
\end{equation}
where $\sigma_\eta(G)$ is the along-scan measurement uncertainty per CCD (e.g. Figure 1 of \citealt{el-badry_generative_2024}) and $\hat{a}_0$ is a \texttt{RUWE}-informed estimate of the photocenter semi-major axis. We then selected targets satisfying 

\begin{equation}
    \label{eq:a0_over_varpi}
    \frac{\hat{a}_0}{\varpi} > 1\,{\rm au}.
\end{equation}

These cuts yielded a total of 292 sources, of which 31 already had an orbital solution reported in the \texttt{gaiadr3.nss\_two\_body\_orbit} catalog. As before, these sources have already been investigated elsewhere; none harbor massive dark companions. 

Luminous companions yield smaller photocenter orbits at fixed mass than dark companions, and so this selection is designed to target binaries with large photocenter orbits corresponding to dark and massive companions. Equation~\ref{eq:a0} will underestimate $a_0$ in cases where the period significantly exceeds the 1000-d DR3 observing baseline, so we expect most binaries selected in this way to have periods of order 1000 d. Given that the binaries' periods are not known a priori and there is significant scatter in the relation between \texttt{RUWE} and $a_0$ \citep{lam_fast_2024, el-badry_inflation_2025}, it is inevitable that some luminous binaries with smaller orbits will enter the sample as well. 

We measured at least one RV for 103 of the remaining 261 candidates. We have thus observed $\approx 39$\% of the remaining 261 objects in the $[\rm M/H]<-1.0$, large $a_0$ sample, or $\approx46$\% of 292 when we include the objects with orbital solutions. Of the 261 candidates lacking orbital solutions, 4 have acceleration or variable acceleration solutions reported in the \texttt{gaiadr3.nss\_acceleration\_astro} catalog. 2 of these sources are in our observed sample.

\subsection{Summary of the sample}

\begin{figure*}
    \centering
    \includegraphics[width=\textwidth]{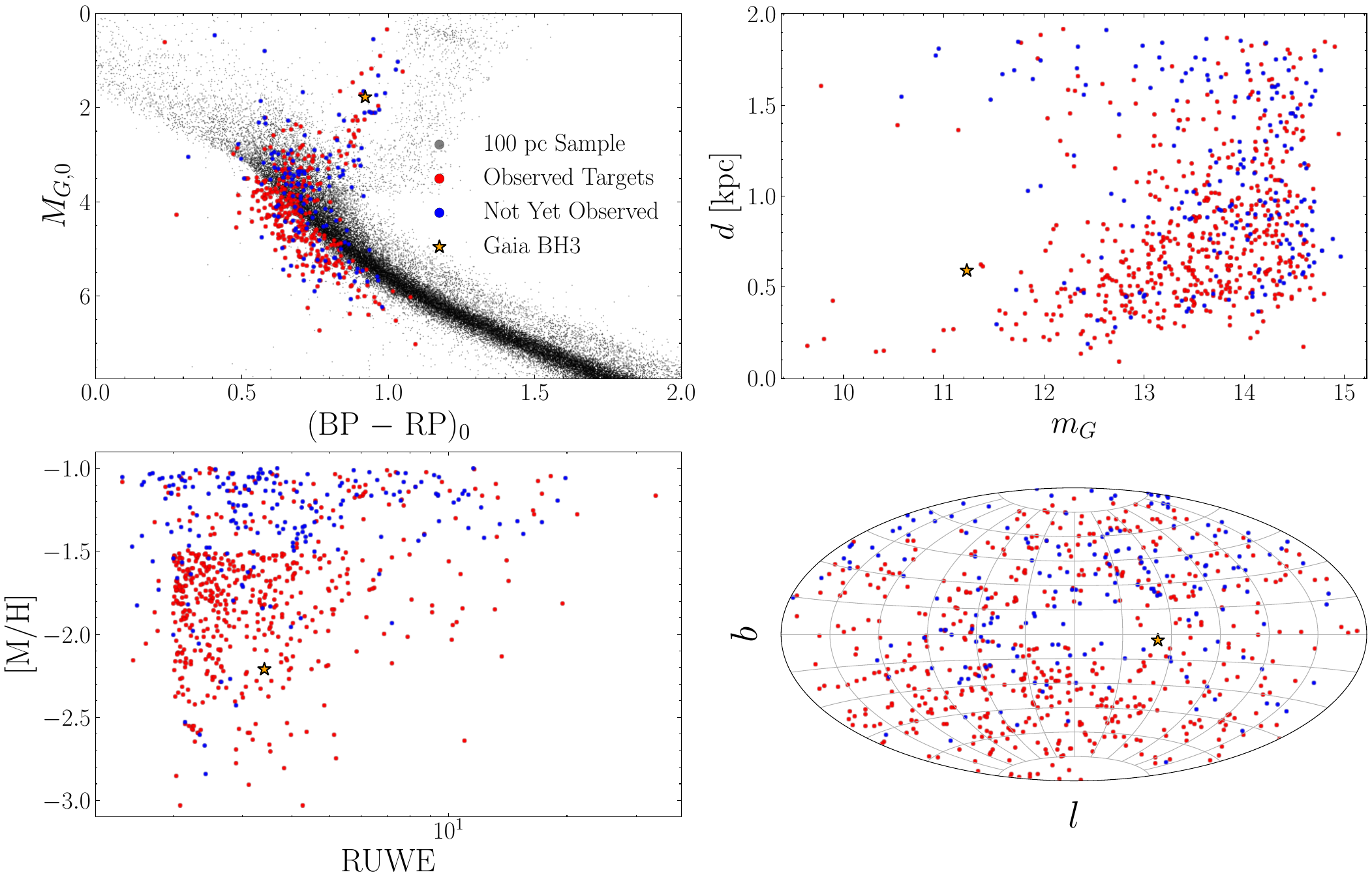}
    \caption{Properties of our sample of low-metallicity stars with elevated \texttt{RUWE}, with Gaia BH3 shown for comparison. Upper left: Color-magnitude diagram. As expected, the low-metallicity stars are bluer than their counterparts in the 100 pc random comparison sample. Upper right: Distance versus apparent magnitude for all targets. They are bright and have mean RVs in DR3. Lower left: Metallicity versus \texttt{RUWE} for all targets. They have [M/H] $< -1$ and high \texttt{RUWE} values indicating binarity. Lower right: Sky distribution of all targets. They are distributed approximately randomly across the sky.}
    \label{fig:sample_properties}
\end{figure*}

Basic properties of our sample of low-metallicity stars with elevated \texttt{RUWE} are summarized in Figure~\ref{fig:sample_properties}. We exclude candidates with orbital solutions. In the upper left panel, we show an extinction-corrected color-magnitude diagram, with a random comparison sample of stars within 100 pc for reference. The extinctions are calculated based on reddenings retrieved from the 3D dust map of \citet{green_2019}. As expected, the low-metallicity stars are bluer than the comparison sample. In the upper right panel, we plot distance versus apparent magnitude for all targets. Our targets tend to lie between $G = 12$--$15$ and have typical distances of $0.5$--$2$ kpc. In the lower left panel, we plot metallicity versus \texttt{RUWE}. All observed targets have both [M/H] $< -1$ and high \texttt{RUWE} values $>1.4$ indicating binarity. The structure in this panel arises because we use different sample selection criteria for binaries with $\rm [M/H] < -1.5$ and $-1.5 < \rm [M/H] < -1.0$. Finally, in the lower right, we show the sky distribution of all targets. The targets are distributed approximately randomly, as expected for halo stars. Some structure is evident, perhaps due in part to our cut of $v_\perp > 70$ km s$^{-1}$, which excludes stars on disk-like orbits.

Our selection cuts depend on the availability of robustly measured parallaxes, but DR3 parallax uncertainties for sources with elevated \texttt{RUWE} are underestimated by a factor of $1$--$4$ \citep{el-badry_inflation_2025}. After inflating the parallax uncertainties of our targets based on the prescription of \citet{el-badry_inflation_2025}, we find that $\approx68\%$ still have parallax to parallax uncertainty ratios $> 5$, implying that most of our candidates still have reasonably well-constraned distances. 

In summary, we obtain at least one epoch RV for a majority of low-metallicity stars with $\rm [M/H] < -1.5$ and \texttt{RUWE} $> 2.0$. We also obtain epoch RVs for an additional subset of sources with $\rm [M/H] < -1.0$, \texttt{RUWE} $> 1.4$, and large photocenter orbits. Most of the $500+$ binaries in our observed sample are expected to have orbital periods $\gtrsim 10^3$ days, though some are likely shorter-period binaries or higher-order multiples that failed one or more of the quality cuts imposed on DR3 astrometric solutions. DR3 reports mean RVs for stars with $G_{\text{RVS}} \lesssim 14$ and RV variability information for stars with $G_{\text{RVS}} \lesssim 12$. For the typical colors of targets in our sample, these correspond roughly to $G < 15$ and $G < 13$, respectively. A complimentary sample of low-metallicity stars with potential BH companions will be investigated in Lam et al. (forthcoming).

\section{Data} 
\label{sec:data}

\subsection{Spectroscopic Follow-up}

We obtained 657 spectra for 528 targets, $\sim 7$ years after the median epoch of the time series data in \textit{Gaia} DR3. Of these spectra, 499 were obtained using FEROS, and 158 were obtained using APF. We used a typical exposure time of 1800s for targets with $G = 14$, and shorter and longer exposure times for brighter and fainter objects. We make all our spectra publicly available at \texttt{https://doi.org/10.5281/zenodo.16888415}.

\subsubsection{FEROS}

We used the Fiberfed Extended Range Optical Spectrograph \citep[FEROS;][]{feros_1999} on the 2.2m ESO/MPG telescope at La Silla Observatory for targets at $\delta < 20^{\circ}$ (programs 113.26XB.001, 114.27SS.001, and 115.28KE.001). The resulting spectra have resolution $R \approx 50,000$ over a spectral range of $350$--$920$ nm, and a typical SNR of 4 at 5178 \AA. We reduced the data using the CERES pipeline \citep{Brahm2017}. The pipeline performs bias subtraction, flat fielding, wavelength calibration, and optimal extraction, and measures and corrects for small shifts in the wavelength solution over the course of a night using simultaneous observations of a ThAr lamp with a second fiber. For each spectrum, we estimated the continuum using a median filter, and then performed continuum normalization.

\subsubsection{APF}

We used the Levy spectrometer on the 2.4m Automated Planet Finder \citep[APF;][]{radovan_apf_2010, apf_2014} telescope at Lick Observatory for targets at $\delta \geq 20^{\circ}$. We used a 2'' slit, yielding spectra with resolution $R \approx 80,000$ over a wavelength range of $373$--$1020$ nm and a typical SNR of 8 at 5178 \AA. We reduced the data using the California Planet Search pipeline \citep{howard_cps_2010, fulton_cps_2015}. We measured and corrected for small shifts in the wavelength solution over the course of a night using telluric absorption lines (see \citealt{nagarajan_spectroscopic_2023} for a description of the method). As with the FEROS spectra, we estimated the continuum of each spectrum using a median filter, and then performed continuum normalization. 

\subsection{Radial Velocities}

\begin{figure*}
    \centering
    \includegraphics[width=\textwidth]{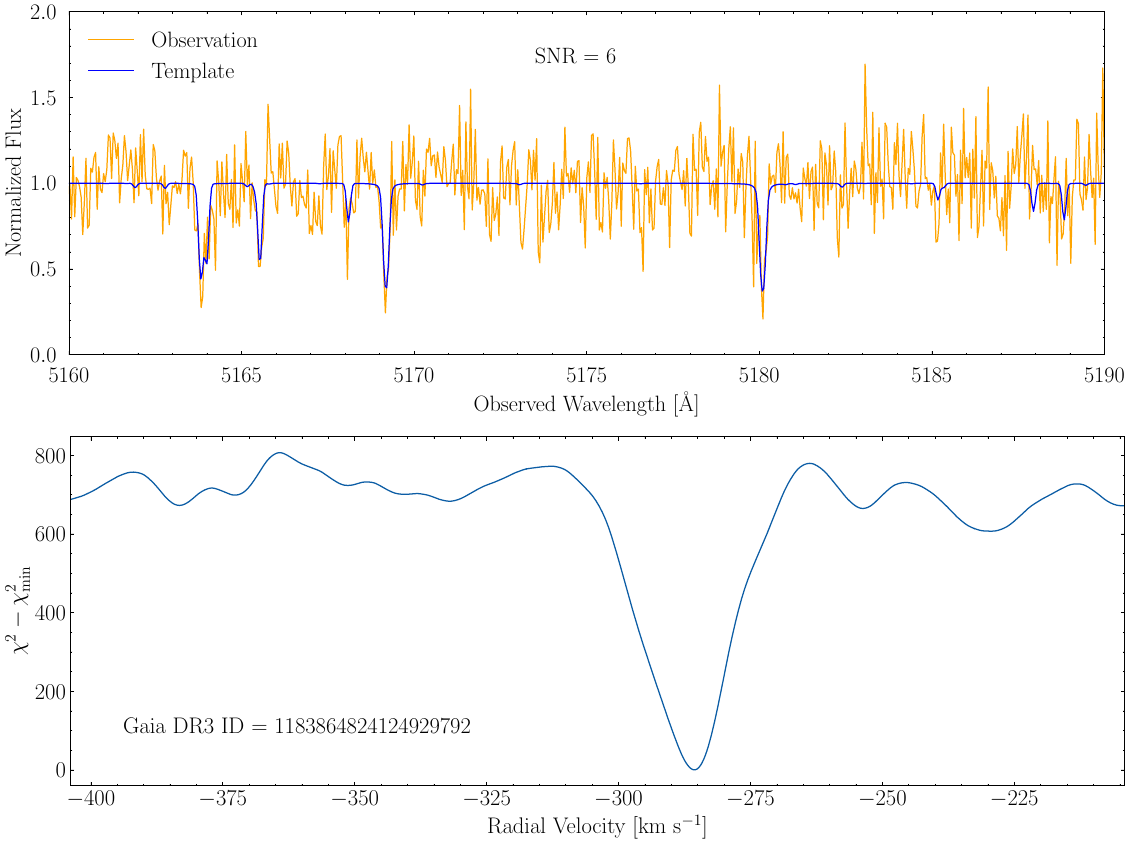}
    \caption{Top: Comparison of a template to a FEROS spectrum of a typical target in our sample. The three strongest lines are due to the Mg I b triplet. Bottom: Chi-squared statistic as a function of RV shift for the same target. The minimum corresponds to the derived RV. We measure a robust RV of $-286.16 \pm 0.45$ km s$^{-1}$ despite the relatively low SNR of 6.}
    \label{fig:good_chi_squared}
\end{figure*}

\begin{figure*}
    \centering
    \includegraphics[width=\textwidth]{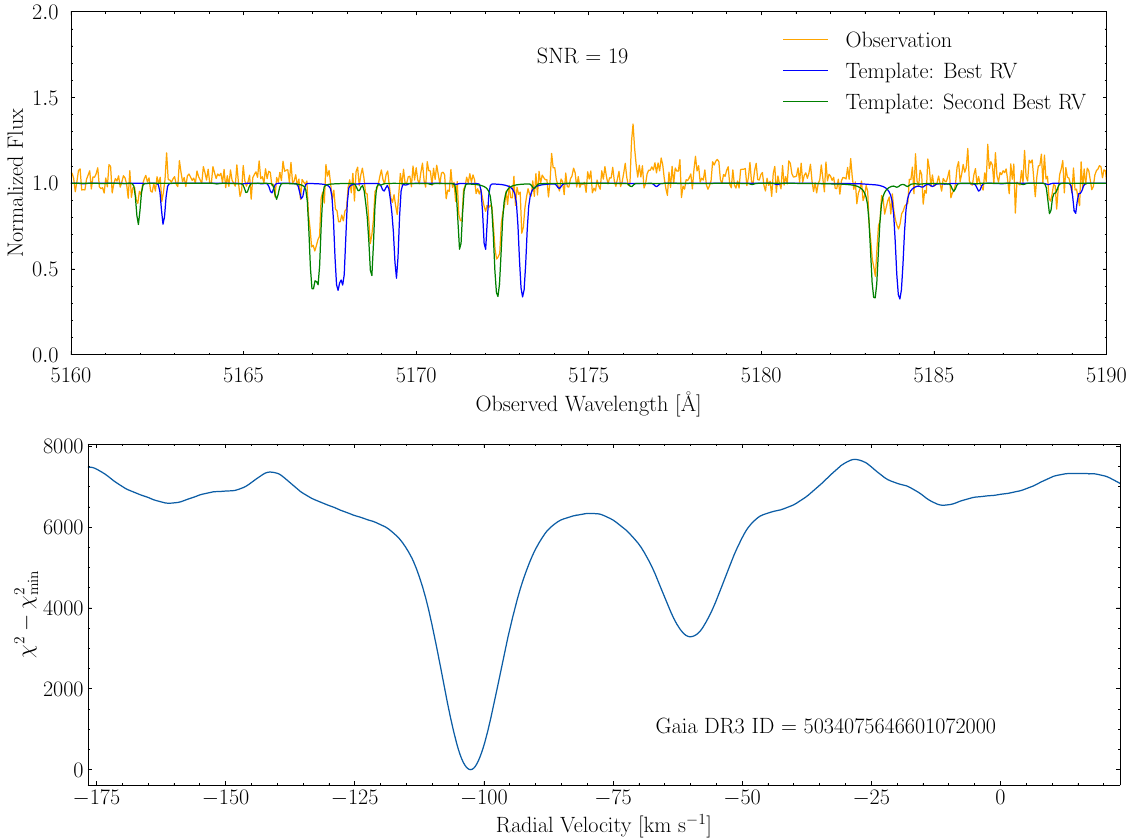}
    \caption{Same as Figure~\ref{fig:good_chi_squared}, but for a target deemed to have a bad template fit. The chi-squared statistic curve is multimodal, suggesting that the system is a double-lined spectroscopic binary (SB2). We show the shifted template spectrum in the lab frame for both plausible RV minima. Lines due to two luminous stars are clearly visible.}
    \label{fig:bad_chi_squared}
\end{figure*}

We measured RVs via comparison with a synthetic template spectrum. Specifically, we measured RVs based on two consecutive orders that display prominent absorption features. We used orders 9 and 10 for FEROS and orders 34 and 35 for APF, corresponding to wavelength regions of 5048--5362\,\AA\, and 5137--5279\,\AA, respectively. These wavelength regions include the Mg I b triplet, which in most of our targets is among the strongest features in the optical. We experimented with a variety of spectral regions and found these to be the most robust; the Balmer lines are too broad, and the Mg I lines are among the strongest lines in metal-poor stars. Many other orders do not contain detectable lines in most of our targets.

We calculate the chi-squared statistic as a function of RV shift, defined as:

\begin{equation}
    \chi^2 = \sum_{\lambda} \left(\frac{F_{\lambda, \text{obs}} - F_{\lambda, \text{template}}}{\sigma_{F_{\lambda, \text{obs}}}}\right)^2,
\end{equation}

\noindent where $F_{\lambda, \text{obs}}$ is the observed flux, $F_{\lambda, \text{template}}$ is the template flux, shifted to the appropriate RV, and $\sigma_{F_{\lambda, \text{obs}}}$ is the uncertainty in the observed flux at wavelength $\lambda$. 

For each target, we chose the template spectrum as the BOSZ Kurucz \citep{Kurucz1979, Kurucz1993, bosz_2024} spectrum with stellar parameters closest to those derived by \citet{andrae_2023}. The template spectra cover a grid of metallicity $\rm [M/H] \in (-2.5, -1.0)$ spaced by $-0.25$, effective temperature $T_{\text{eff}} \in (4000 \text{ K}, 6500 \text{ K})$ spaced by $250$ K, and surface gravity $\log{\left(g / \text{cm s$^{-2}$}\right)} \in (2.5 , 4.5)$ spaced by $0.5$. The template spectra are at a resolution of $R = 50,000$ and assume no rotational broadening.

The best-fit RV is the shift for which the $\chi^2$ statistic is minimized. Uncertainties on RVs are derived based on the width of the minimum, i.e., the difference at which the $\chi^2$ statistic changes by one. All of our RVs are reported in Table~\ref{tab:comparison} in Appendix~\ref{sec:all_rvs}.

We manually inspected the template fit in all cases where the epoch RV differed from the mean DR3 RV by more than 10 km s$^{-1}$ (see Table~\ref{tab:interesting}). We do not account for the error in the mean DR3 RV when adopting this threshold, since that error often arises from orbital motion rather than statistical uncertainty. During this vetting process, we visually identified a few objects where a joint fit to both orders produced a less accurate RV than a fit to only one of the orders. In these cases, we adopt the more accurate RV derived from the best fit to one of the orders alone.

We show the observed FEROS spectrum for a typical target, along with the template spectrum used to derive the corresponding best-fit RV, in Figure~\ref{fig:good_chi_squared}. We plot a spectral region centered on the Mg I b triplet, from which most of the signal used to derive the RV arises. We also show the $\chi^2$ statistic as a function of RV shift. The location of the minimum in this curve corresponds to the best-fit RV.

\subsection{Flagged Spectra}
\label{sec:problematic}

We show the observed FEROS spectrum and $\chi^2$ statistic as a function of RV shift for a target for which the derived RV is unreliable in Figure~\ref{fig:bad_chi_squared}. The $\chi^2$ statistic curve clearly has two local minima, suggesting that the system is a double-lined spectroscopic binary (i.e., SB2). Visually, the single-star template shifted into the lab frame does not seem to be a good fit to the observed spectrum for either plausible RV.

Some spectra have low signal-to-noise values (i.e., SNR $< 2$ at 5200\,\AA) or are likely contaminated with moonlight (i.e. have barycenter correction within $10$ km s$^{-1}$ of the derived RV). Other spectra do not have prominent absorption lines due to high effective temperatures (i.e., the source is a hot star) or rapid rotation (i.e., the source has a high $v \sin i$). In all these cases, we flag the derived RV as unreliable, remove the target from our sample of promising candidates, and report the flag in Table~\ref{tab:comparison}.

\section{Results} 
\label{sec:analysis}

\begin{figure*}[h!]
    \centering
    \includegraphics[width=\textwidth]{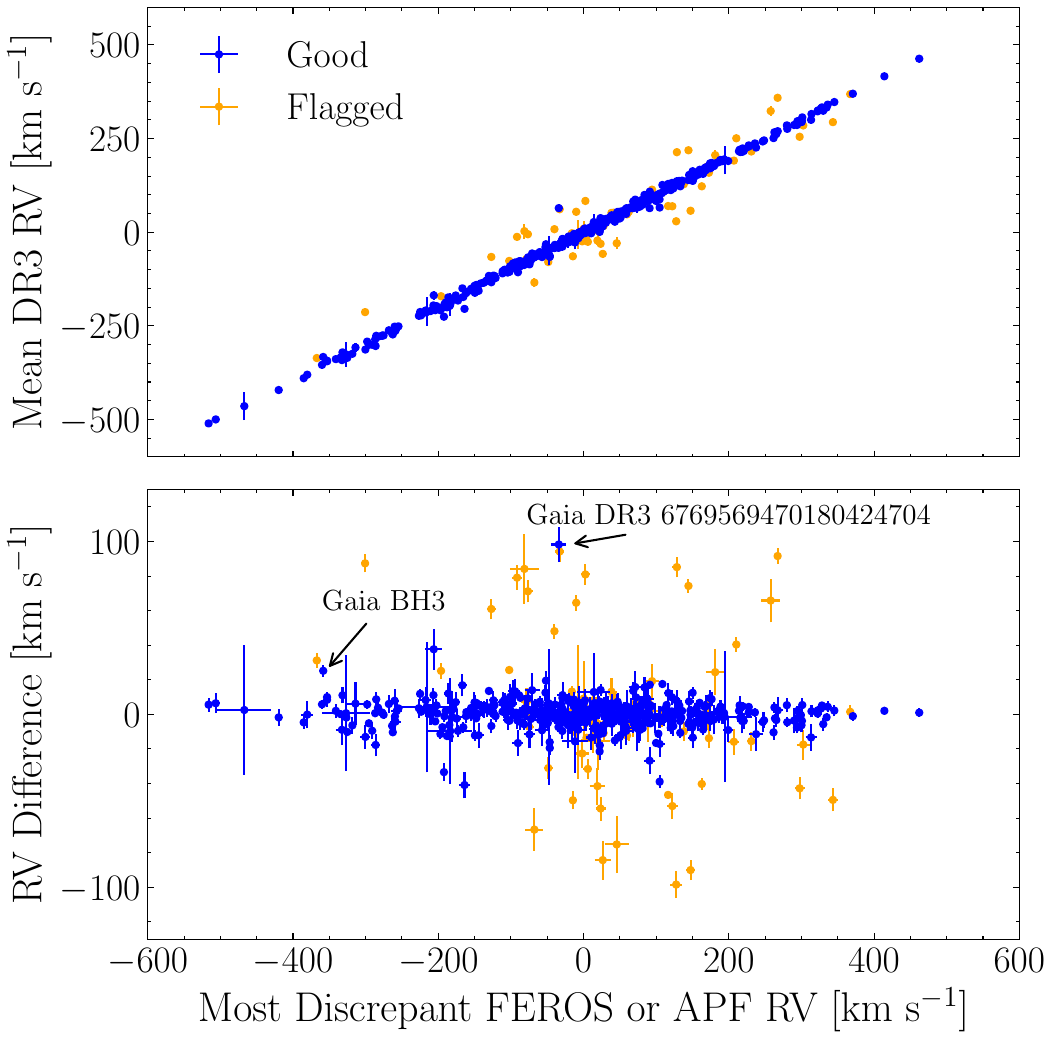}
    \caption{Comparison of our derived RVs to the mean RV reported in DR3. RVs for flagged spectra (see Section~\ref{sec:problematic}) are plotted in orange. In most cases, the RVs are in good agreement. In several cases, the RVs are discrepant, suggesting the presence of a faint or dark companion. We highlight both Gaia BH3 and \textit{Gaia} DR3 6769569470180424704, which has the largest RV discrepancy in our sample.}
    \label{fig:compare_gaia_rvs}
\end{figure*}

\begin{figure*}
    \centering
    \includegraphics[width=\textwidth]{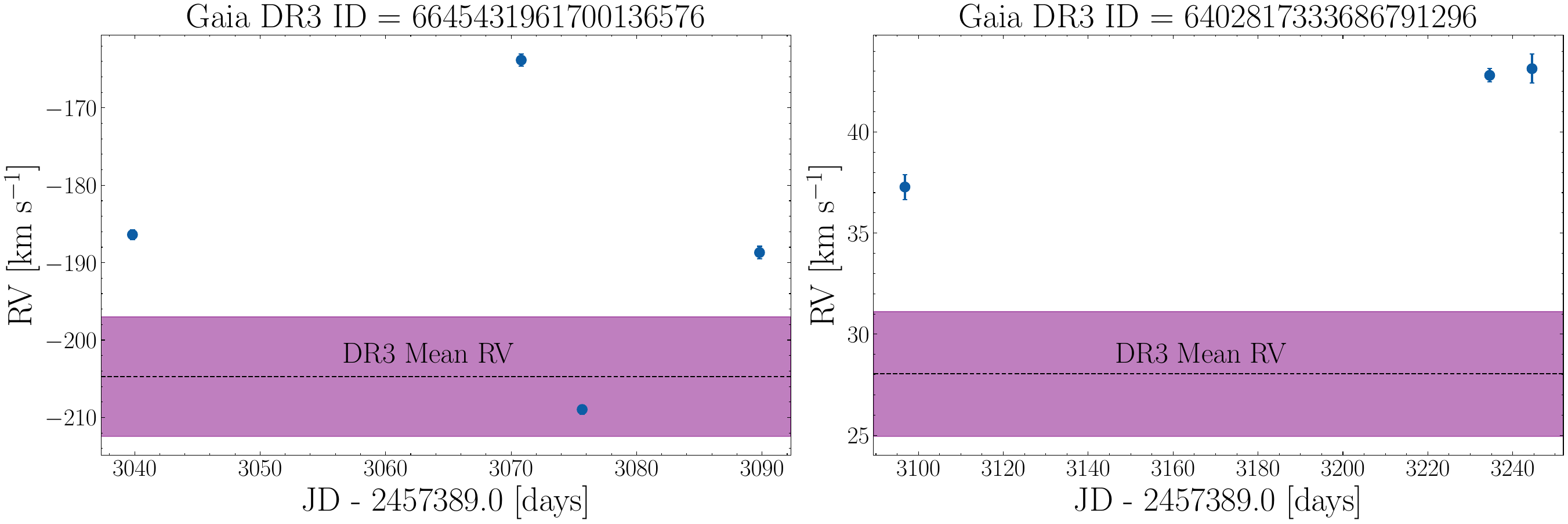}
    \caption{Examples of interesting targets with multiple RV measurements. In the left panel, we see RV variability on short timescales, suggesting that the system is potentially a hierarchical triple (see Section \ref{sec:multiple}). In the right panel, we see evidence of slower RV variability, but extended RV monitoring is necessary to confirm or rule out a compact object companion.}
    \label{fig:multi_rv_figure}
\end{figure*}

We compare our derived RVs to the mean RVs reported in DR3 in the top panel of Figure~\ref{fig:compare_gaia_rvs}. We plot good RVs in blue, and RVs for spectra that were flagged in orange. We show a residual plot in the bottom panel. In most cases, the epoch RVs are in good agreement with the mean RVs. In several cases, the two RVs are discrepant beyond the statistical uncertainties, suggesting the possibility of a hidden compact object companion.

In Gaia BH3, the star’s RV is at least 20 km s$^{-1}$ and 10 km s$^{-1}$ different from the mean RV published in \text{Gaia} DR3 over $\approx70$\% and $\approx85$\% of the orbit, respectively. Since the instantaneously measured RV of a given target is expected to be significantly different from its mean RV over a large portion of its orbit (Figure~\ref{fig:bh3_rv_astro}), a large RV shift is potentially indicative of a massive companion. Conversely, a small RV shift relative to the DR3 mean disfavors a massive companion, though it does not rule one out. We explore this quantitatively in Section~\ref{sec:discussion}.

We list all promising (i.e., non-flagged) targets with RV discrepancy $> 10$ km s$^{-1}$ in Table~\ref{tab:interesting}. We have manually vetted each candidate in this table, removing targets that are luminous binaries, hot stars, or fast rotators. Of the 67 interesting candidates that remain, 58 are from the $\rm [M/H] < -1.5$, \texttt{RUWE} $> 2.0$ sample, while 9 are from the $\rm [M/H] < -1.0$, \texttt{RUWE} $> 1.4$, large $a_0$ sample. We verify that our search method successfully recovers Gaia BH3, which has \text{Gaia} DR3 source ID 4318465066420528000. Our search returns 6 non-flagged candidates with maximum RV discrepancies larger than Gaia BH3. 

If we account for the uncertainties on our RV measurements and the reported \textit{Gaia} DR3 RVs, only 49 of the 67 interesting candidates listed in Table~\ref{tab:interesting} have RV discrepancies $> 0$ at the $2\sigma$ level. However, these uncertainties are dominated by the uncertainty in the \textit{Gaia} DR3 mean RV, which is often a result of orbital motion rather than the statistical uncertainty on an epoch RV measurement (see Section~\ref{sec:discussion} for further details on the process by which DR3 RVs are computed). As such, we still include the 18 candidates whose largest RV shift is statistically consistent with zero to within $2\sigma$, as we do not think that this substantially undermines the evidence indicating the presence of a potential compact object companion.

Of the flagged objects plotted in orange in Figure~\ref{fig:compare_gaia_rvs}, 52 correspond to low-SNR spectra, while 4 have spectra that are potentially contaminated with moonlight. The remainder were flagged after manual vetting, with 11, 8, and 2 objects determined to be double-lined spectroscopic binaries, hot stars, and fast rotators, respectively. Many of the flagged objects have large RV discrepancies that we find to be spurious, and many of the candidates with large RV discrepancies are flagged. However, several genuine candidates survive the vetting process, and are listed in Table~\ref{tab:interesting}.

\subsection{Objects with Multiple RVs}
\label{sec:multiple}

\begin{figure}
    \centering
    \includegraphics[width=\columnwidth]{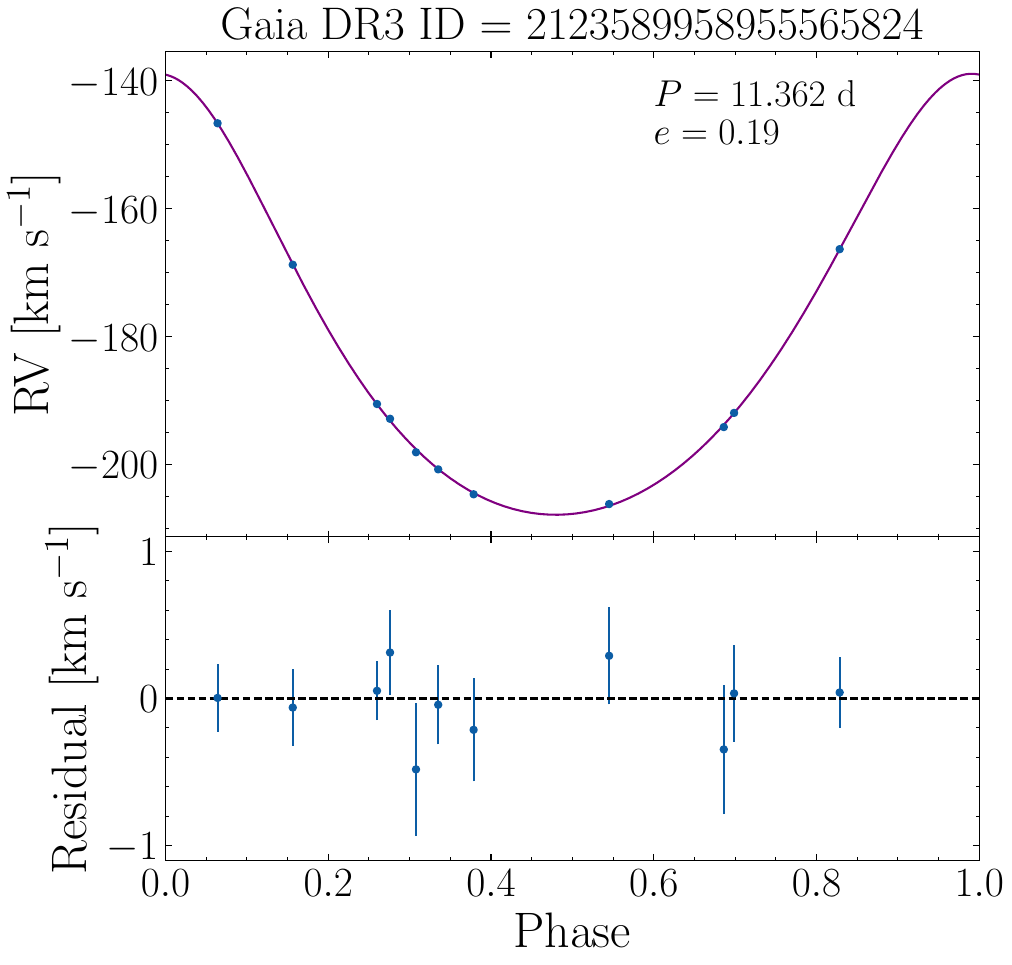}
    \caption{Best-fit RV curve and residuals for \textit{Gaia} DR3 2123589958955565824. We measured RVs at 11 epochs using APF. The minimum companion mass is only $\approx 0.4\,M_{\odot}$, implying that the system is unlikely to host a BH. While the observed DR3 \texttt{RUWE} is $2.45$, the expected DR3 \texttt{RUWE} for the fitted orbit is $\approx 1$, rendering it likely that the source is a hierarchical triple.}
    \label{fig:triple_rv_figure}
\end{figure}

We are attempting to obtain at least one additional follow-up RV for each target with RV discrepancy $> 10$ km~s$^{-1}$. We plot RVs over time for two typical targets in Table~\ref{tab:interesting} with RV discrepancy $> 10$ km s$^{-1}$ in Figure~\ref{fig:multi_rv_figure}. The target in the left panel displays RV variability on short timescales. This suggests that the source is an unresolved hierarchical triple; in this scenario, the observed RV variability arises from the inner binary, while the large \texttt{RUWE} value is due to the motion of the outer orbit. The target in the right panel displays slower RV variability that suggests a long orbital period, consistent with our expectations for typical binaries in our sample.

For one target in our sample, \textit{Gaia} DR3 2123589958955565824, we measured 11 RVs, which is enough to constrain the full orbital solution.  We use ensemble MCMC sampling \citep[\texttt{emcee};][]{Foreman-Mackey_2013} with 64 walkers and 10000 iterations to fit the RVs with a two-body Keplerian orbit and show the best-fit RV curve in Figure~\ref{fig:triple_rv_figure}. We find a short orbital period of $P = 11.362 \pm 0.001$ d and a moderate eccentricity of $e = 0.19 \pm 0.01$. The best-fit RV semi-amplitude of $34.5 \pm 0.3$ km s$^{-1}$ leads to a mass function of $\approx 0.046\,M_{\odot}$ and a minimum companion mass of $\approx 0.39\,M_{\odot}$, assuming a luminous star mass of $0.75\,M_{\odot}$. This makes the system unlikely to contain a BH. Using \texttt{gaiamock}, we find that the predicted DR3 \texttt{RUWE} for the derived orbit and typical inclinations is $\approx 1$, much smaller than the observed DR3 \texttt{RUWE} of $2.45$. We conclude that the system is likely a hierarchical triple, with the RVs tracing the motion of the inner binary.

While the total number of RV epochs per target ranges from $2$--$20$, and we observe some RV variability in all targets, we have not yet fully solved the orbit of any of our other targets. For targets with orbital periods $\gtrsim 1000$ days, a few RVs taken over a short time frame are insufficient to fit a full orbital solution and tightly constrain the companion mass. Since extended RV monitoring is necessary to determine the orbital period and confirm or rule out a BH or NS companion, we will continue following up interesting candidates over the next several months. Upon release of DR4, we will derive constraints on the companion masses by jointly fitting our long-baseline measurements with epoch \textit{Gaia} astrometry.

\startlongtable
\begin{deluxetable*}{ccccc}
\tablecaption{Derived radial velocities (RVs) for all promising candidates with RV difference between epoch RV and mean DR3 RV $> 10$ km s$^{-1}$. \label{tab:interesting}}
\tablehead{\colhead{\textit{Gaia} DR3 ID} & \colhead{$N_{\text{RVs, good}}$} & \colhead{Most Discrepant FEROS or APF RV} &  \colhead{\textit{Gaia} DR3 Mean RV} & \colhead{Largest RV Shift} \\
& & km s$^{-1}$ & km s$^{-1}$ & km s$^{-1}$ \\
\colhead{(1)} & \colhead{(2)} & \colhead{(3)} & \colhead{(4)} & \colhead{(5)}}
\startdata
6769569470180424704 & 20 & $-34.02 \pm 1.62$ & $64.26 \pm 10.26$ & $98.28 \pm 10.39$ \\
6645431961700136576 & 4 & $-163.82 \pm 0.77$ & $-204.71 \pm 7.70$ & $40.89 \pm 7.74$ \\
6393888749512565760 & 2 & $104.85 \pm 0.09$ & $65.91 \pm 3.57$ & $38.94 \pm 3.57$ \\
2123589958955565824 & 11 & $-206.18 \pm 0.33$ & $-168.51 \pm 11.93$ & $37.67 \pm 11.94$ \\
1109010106866929536 & 8 & $-192.15 \pm 0.38$ & $-225.58 \pm 5.24$ & $33.43 \pm 5.25$ \\
6714154840051650560 & 1 & $91.02 \pm 0.41$ & $64.23 \pm 7.79$ & $26.79 \pm 7.80$ \\
4318465066420528000 & 2 & $-358.38 \pm 0.03$ & $-333.18 \pm 3.38$ & $25.20 \pm 3.38$ \\
5044327218140737536 & 1 & $182.51 \pm 0.65$ & $205.68 \pm 13.26$ & $23.17 \pm 13.27$ \\
4466423837863123328 & 3 & $-313.02 \pm 0.63$ & $-336.00 \pm 4.36$ & $22.98 \pm 4.41$ \\
6463410835377385472 & 2 & $22.19 \pm 0.55$ & $0.77 \pm 5.02$ & $21.42 \pm 5.05$ \\
5817399066319766912 & 2 & $-51.73 \pm 0.92$ & $-32.23 \pm 5.38$ & $19.50 \pm 5.46$ \\
3190523740098447744 & 2 & $-46.45 \pm 0.05$ & $-65.92 \pm 4.26$ & $19.47 \pm 4.26$ \\
3252546886080448384 & 1 & $22.43 \pm 0.14$ & $4.52 \pm 3.48$ & $17.91 \pm 3.48$ \\
1183864824124929792 & 2 & $-286.18 \pm 0.46$ & $-303.88 \pm 6.54$ & $17.70 \pm 6.56$ \\
5767212132939510400 & 3 & $108.62 \pm 0.23$ & $126.17 \pm 1.38$ & $17.55 \pm 1.40$ \\
5470326985368066304 & 1 & $104.71 \pm 0.72$ & $87.58 \pm 7.48$ & $17.13 \pm 7.51$ \\
2474847090389656064 & 3 & $91.26 \pm 0.20$ & $108.36 \pm 1.61$ & $17.10 \pm 1.63$ \\
4449971364341398272 & 3 & $-166.72 \pm 0.42$ & $-149.89 \pm 6.44$ & $16.83 \pm 6.45$ \\
6820303307501354496 & 1 & $-90.31 \pm 0.78$ & $-106.92 \pm 7.46$ & $16.61 \pm 7.50$ \\
2788486767941706624 & 3 & $-229.91 \pm 0.53$ & $-213.35 \pm 5.21$ & $16.56 \pm 5.24$ \\
677442515834924416 & 3 & $99.91 \pm 0.21$ & $83.36 \pm 2.10$ & $16.55 \pm 2.11$ \\
6177352097769902208 & 2 & $-46.58 \pm 0.26$ & $-62.66 \pm 4.00$ & $16.08 \pm 4.01$ \\
1589175287211605376 & 1 & $71.24 \pm 0.38$ & $87.06 \pm 9.09$ & $15.82 \pm 9.10$ \\
6822730754297502720 & 2 & $-11.79 \pm 0.95$ & $-27.42 \pm 18.27$ & $15.63 \pm 18.30$ \\
3529854748080356352 & 1 & $84.01 \pm 0.05$ & $99.61 \pm 5.97$ & $15.60 \pm 5.97$ \\
3195269030422383488 & 1 & $13.74 \pm 0.93$ & $29.14 \pm 7.83$ & $15.40 \pm 7.88$ \\
6402817333686791296 & 3 & $43.13 \pm 0.72$ & $28.04 \pm 3.08$ & $15.09 \pm 3.16$ \\
2402811860245032448 & 3 & $23.50 \pm 0.03$ & $37.60 \pm 3.30$ & $14.10 \pm 3.30$ \\
2429303635139814400 & 1 & $-70.78 \pm 0.66$ & $-56.80 \pm 10.18$ & $13.98 \pm 10.20$ \\
1294817813795066880 & 1 & $68.46 \pm 0.51$ & $82.04 \pm 5.29$ & $13.58 \pm 5.31$ \\
1694181735744784768 & 1 & $-130.24 \pm 0.11$ & $-116.69 \pm 2.53$ & $13.55 \pm 2.53$ \\
2619820000613684992 & 1 & $-44.45 \pm 0.68$ & $-30.95 \pm 6.95$ & $13.50 \pm 6.98$ \\
5510992972680488832 & 1 & $150.43 \pm 0.65$ & $136.99 \pm 6.07$ & $13.44 \pm 6.11$ \\
2438001734187492736 & 2 & $-97.24 \pm 0.18$ & $-83.83 \pm 6.20$ & $13.41 \pm 6.20$ \\
6689320926830067328 & 1 & $10.56 \pm 0.30$ & $-2.82 \pm 7.10$ & $13.38 \pm 7.11$ \\
6646169837081308032 & 1 & $313.32 \pm 0.55$ & $299.98 \pm 7.59$ & $13.34 \pm 7.61$ \\
5889520363370193536 & 1 & $-94.12 \pm 0.44$ & $-80.83 \pm 7.11$ & $13.29 \pm 7.13$ \\
4324114150961924992 & 2 & $49.69 \pm 0.38$ & $36.46 \pm 1.03$ & $13.23 \pm 1.10$ \\
4036323751356613120 & 2 & $-300.32 \pm 0.45$ & $-313.37 \pm 6.79$ & $13.05 \pm 6.81$ \\
1445952348525801344 & 1 & $14.15 \pm 0.08$ & $27.08 \pm 22.82$ & $12.93 \pm 22.82$ \\
1363047492460703744 & 1 & $-184.23 \pm 0.29$ & $-196.91 \pm 4.51$ & $12.68 \pm 4.52$ \\
1028448477581901184 & 1 & $-52.40 \pm 0.14$ & $-39.82 \pm 4.59$ & $12.58 \pm 4.59$ \\
4404085239419591552 & 2 & $-187.95 \pm 0.39$ & $-200.46 \pm 2.46$ & $12.51 \pm 2.49$ \\
6657756868572776832 & 2 & $150.28 \pm 0.06$ & $162.76 \pm 3.37$ & $12.48 \pm 3.37$ \\
3838404232257203072 & 1 & $116.84 \pm 0.53$ & $129.14 \pm 5.89$ & $12.30 \pm 5.91$ \\
2281492851046104064 & 1 & $-149.54 \pm 0.18$ & $-161.73 \pm 2.96$ & $12.19 \pm 2.96$ \\
1805460872350903168 & 1 & $-144.36 \pm 0.54$ & $-156.45 \pm 7.24$ & $12.09 \pm 7.26$ \\
1911515915671512064 & 1 & $-186.06 \pm 0.29$ & $-173.99 \pm 6.13$ & $12.07 \pm 6.14$ \\
3619929278952732928 & 1 & $51.08 \pm 0.69$ & $39.05 \pm 8.07$ & $12.03 \pm 8.10$ \\
4792163228561925504 & 1 & $20.64 \pm 0.67$ & $8.67 \pm 5.15$ & $11.97 \pm 5.20$ \\
2835800093315146496 & 2 & $-224.95 \pm 0.23$ & $-213.09 \pm 3.03$ & $11.86 \pm 3.04$ \\
1343879087778316544 & 1 & $-74.33 \pm 0.27$ & $-85.95 \pm 8.14$ & $11.62 \pm 8.14$ \\
4344582418904624256 & 1 & $237.59 \pm 0.33$ & $226.13 \pm 9.81$ & $11.46 \pm 9.82$ \\
4480417902167849728 & 3 & $-196.98 \pm 0.08$ & $-208.32 \pm 2.86$ & $11.34 \pm 2.87$ \\
5000838062927554176 & 1 & $103.09 \pm 0.24$ & $91.90 \pm 2.39$ & $11.19 \pm 2.40$ \\
4463022464283196928 & 1 & $-206.62 \pm 0.69$ & $-195.49 \pm 4.30$ & $11.13 \pm 4.35$ \\
6288565568615343488 & 1 & $-28.83 \pm 0.48$ & $-17.72 \pm 7.79$ & $11.11 \pm 7.81$ \\
3638399772924914176 & 1 & $-88.54 \pm 0.33$ & $-77.50 \pm 2.25$ & $11.04 \pm 2.27$ \\
1558668134509319040 & 2 & $-331.81 \pm 0.19$ & $-320.78 \pm 4.74$ & $11.03 \pm 4.74$ \\
5059047789053289856 & 1 & $26.60 \pm 0.30$ & $15.65 \pm 1.78$ & $10.95 \pm 1.81$ \\
3270069940330680576 & 2 & $21.49 \pm 0.54$ & $10.78 \pm 5.09$ & $10.71 \pm 5.12$ \\
5186664770791126912 & 1 & $133.24 \pm 0.36$ & $122.56 \pm 2.81$ & $10.68 \pm 2.84$ \\
5300091008802250112 & 1 & $261.66 \pm 1.48$ & $251.25 \pm 4.40$ & $10.41 \pm 4.64$ \\
6903293654893289856 & 1 & $-262.57 \pm 0.07$ & $-272.92 \pm 1.17$ & $10.35 \pm 1.17$ \\
1115628342232842368 & 10 & $-325.25 \pm 0.06$ & $-335.43 \pm 2.27$ & $10.18 \pm 2.27$ \\
1698379915953323008 & 1 & $-101.07 \pm 0.29$ & $-90.96 \pm 7.24$ & $10.11 \pm 7.25$ \\
2292990925172085632 & 1 & $-353.07 \pm 0.15$ & $-343.07 \pm 2.66$ & $10.00 \pm 2.66$ \\
\enddata
\end{deluxetable*}

\subsection{A Promising Candidate}

\begin{figure*}
    \centering
    \includegraphics[width=\textwidth]{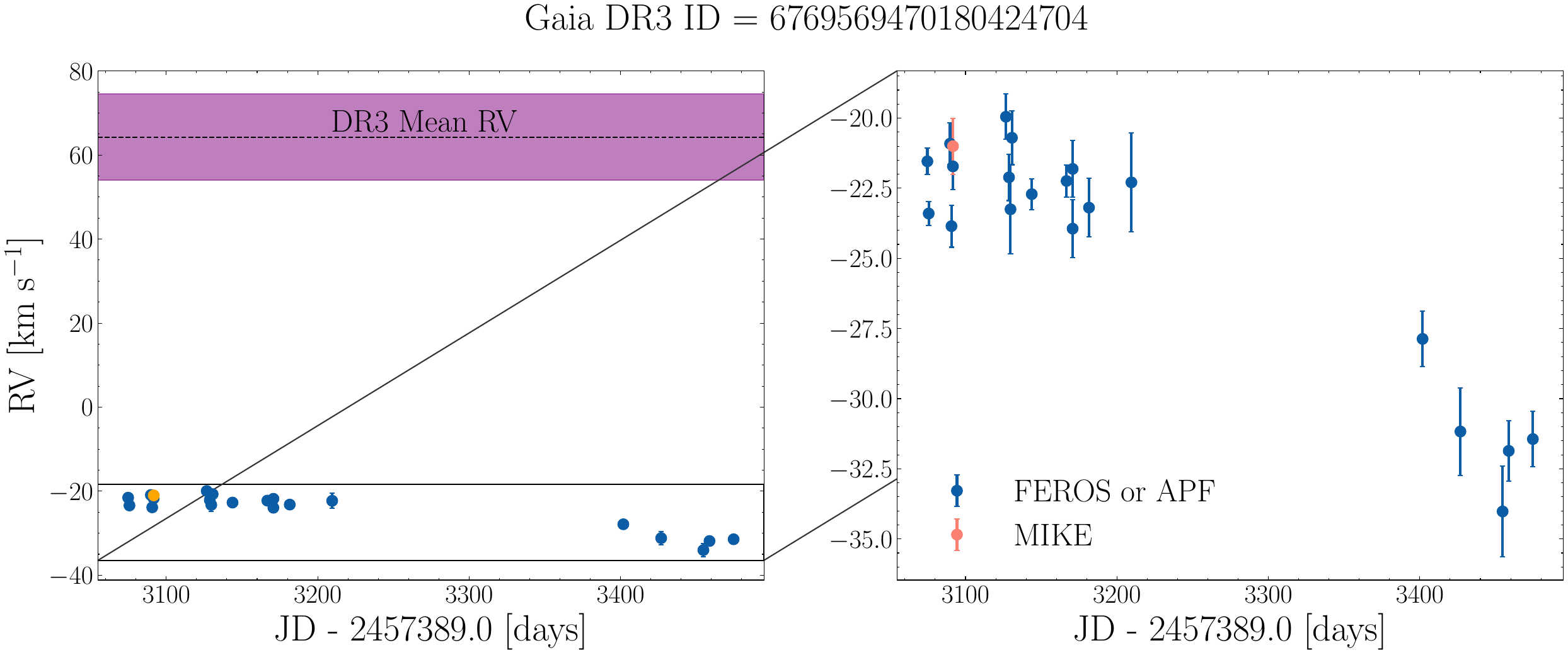}
    \caption{RV measurements over time for \textit{Gaia} DR3 6769569470180424704, the target in our sample with the largest discrepancy between its epoch RVs and the mean RV reported in DR3. The instantaneous RVs measured in 2024--2025 differ from the DR3 mean RV in 2014--2017 by more than 80 km s$^{-1}$. The RVs show evidence of acceleration over a time baseline of a few months, but further RV monitoring is necessary to determine the true nature of this compact object candidate.}
    \label{fig:interesting_object}
\end{figure*}

\begin{figure*}
    \centering
    \includegraphics[width=\textwidth]{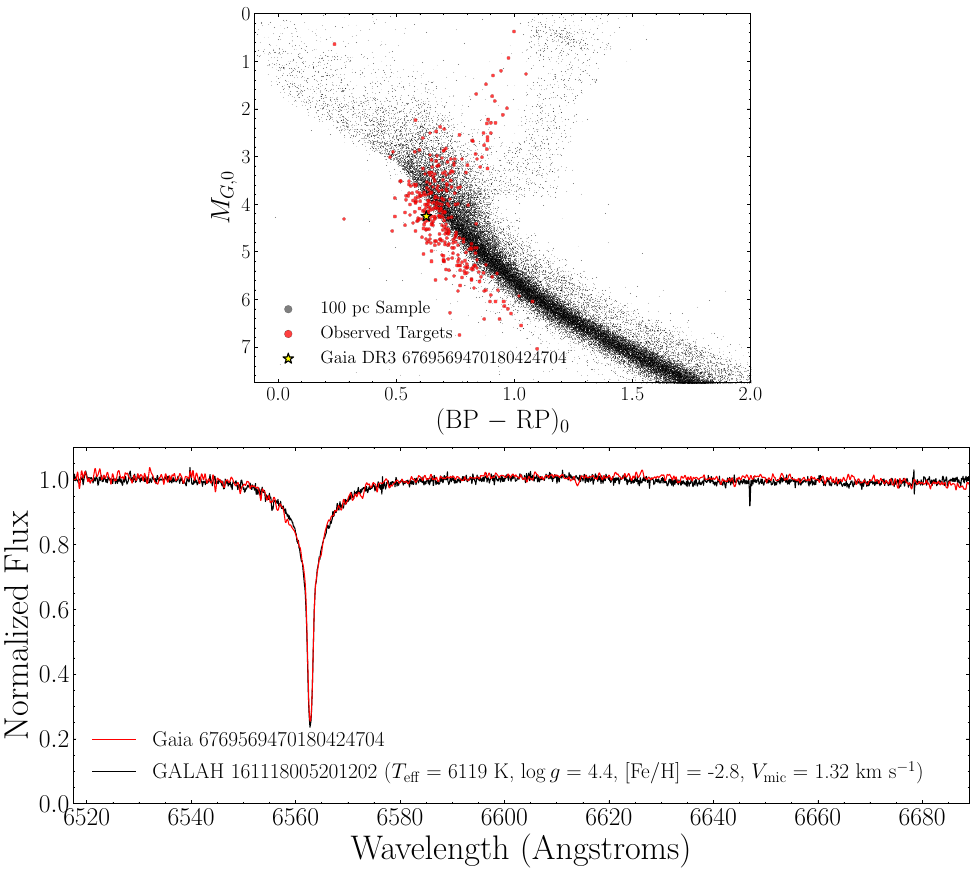}
    \caption{Top: Color-magnitude diagram of \textit{Gaia} DR3 6769569470180424704, the BH candidate with the largest RV discrepancy in our observed sample. The target, which is plotted in yellow, is bluer than the main sequence of the 100 pc random comparison sample, but has typical colors for a metal-poor halo star. Bottom: Cutout of a high-resolution optical spectrum of \textit{Gaia} DR3 6769569470180424704 obtained using the MIKE spectrograph; it appears to be a typical $\sim0.7\,M_{\odot}$ metal-poor main sequence star. The MIKE spectrum of the target is compared against the corresponding region of the spectrum of the object determined to be the closest match in GALAH DR3. The two spectra appear to be quite similar, ruling out the presence of a secondary that makes an appreciable contribution to the total light and has strong H$\alpha$ absorption.}
    \label{fig:galah_comparison}
\end{figure*}

The target \textit{Gaia} DR3 6769569470180424704 has the largest discrepancy ($\approx 98$ km s$^{-1}$) between its epoch RV(s) and its \textit{Gaia} DR3 mean RV in Table~\ref{tab:interesting}. We obtained several follow-up spectra to investigate the true nature of this object. We present all of our epoch RVs in Figure \ref{fig:interesting_object}. We find slow acceleration of $(-11 \pm 2)$ km s$^{-1}$ yr$^{-1}$, with the RV discrepancy growing over time. Extended RV monitoring is warranted.

We obtained a high signal-to-noise spectrum with the Magellan Inamori Kyocera Echelle (MIKE) spectrograph on the Magellan Clay telescope at Las Campanas Observatory \citep{2003SPIE.4841.1694B} on JD 2460480.7435. We used the 0.7'' slit with an exposure time of 2700s, yielding spectral resolution $R \approx 40,000$ over a spectral range of 333--968 nm. The typical SNR per pixel at 5175\,\AA\, is $\approx 70$. We reduced the spectrum using the MIKE pipeline within CarPy \citep{Kelson_2000, Kelson_2003}. To combine the orders, we perform an inverse variance-weighted average, and normalize the resulting spectrum using a polynomial spline fit to continuum wavelengths.

Using the approach described by \citet{reggiani_2022}, we jointly fit the SED and high-resolution spectrum to derive atmospheric parameters for this target. We also measure the equivalent widths of atomic absorption lines to infer elemental abundances. We provide our best-fit parameters and abundances in Table~\ref{tab:mike_params}. The reported uncertainties are statistical, and do not include systematic errors. For comparison, \citet{andrae_2023} estimate $\rm [Fe/H] = -2.3$, $T_{\text{eff}} = 6200$ K, and $\log{\left(g / \text{cm s$^{-2}$}\right)} = 4.2$. The atmospheric parameters we derive are typical for a low-metallicity halo star (given our selection criteria), and are in reasonable agreement with those provided by \citet{andrae_2023}. The target's abundance pattern is unremarkable for a low-metallicity halo star. We verify that non-local thermodynamic equilibrium corrections do not strongly affect any of our derived abundances.

\begin{deluxetable*}{ccc}
\tablecaption{Stellar parameters and abundances of the red giant in \text{Gaia} DR3 6769569470180424704 based on a joint fit of the SED and MIKE spectrum. $N_{\text{lines}}$ is the number of independent absorption lines used to infer the abundance of each element. \label{tab:mike_params}}
\tablehead{\colhead{Parameter} & \colhead{Value} & $N_{\text{lines}}$ \\
\colhead{(1)} & \colhead{(2)} & \colhead{(3)}}
\startdata
$T_{\text{eff}}$ & $6432 \pm 30$ K & \\
$\log{\left(g / \text{cm s$^{-2}$}\right)}$ & $4.42 \pm 0.01$ & \\
$[\text{Fe}/\text{H}]$ & $-2.29 \pm 0.17$ & \\
$v_{\text{micro}}$ & $1.51 \pm 0.19$ km s$^{-1}$ & \\
$t_{\text{age}}$ & $13.0 \pm 0.8$ Gyr & \\
$M_*$ & $0.72 \pm 0.01\,M_{\odot}$ & \\
\hline
$A(\text{Li I})$ & $2.322$ & 1 \\
$\rm[Na\,I/Fe]$ & $0.142 \pm 0.068$ & 2 \\
$\rm[Mg\,I/Fe]$ & $0.269 \pm 0.096$ & 7 \\
$\rm[Al\,I/Fe]$ & $-0.809 \pm 0.111$ & 2 \\
$\rm[Si\,I/Fe]$ & $-0.238 \pm 0.059$ & 1 \\
$\rm[Ca\,I/Fe]$ & $0.263 \pm 0.039$ & 12 \\
$\rm[Sc\,II/Fe]$ & $0.194 \pm 0.090$ & 12 \\
$\rm[Ti\,I/Fe]$ & $0.372 \pm 0.055$ & 5 \\
$\rm[Ti\,II/Fe]$ & $0.285 \pm 0.061$ & 27 \\
$\rm[Cr\,II/Fe]$ & $0.012 \pm 0.090$ & 3 \\
$\rm[Mn\,I/Fe]$ & $-0.508 \pm 0.069$ & 3 \\
$\rm[Fe\,I/H]$ & $-2.311 \pm 0.045$ & 74 \\
$\rm[Fe\,II/H]$ & $-2.289 \pm 0.125$ & 6 \\
$\rm[Co\,I/Fe]$ & $0.226 \pm 0.049$ & 9 \\
$\rm[Ni\,I/Fe]$ & $-0.013 \pm 0.080$ & 13 \\
$\rm[Sr\,II/Fe]$ & $-0.169 \pm 0.104$ & 2 \\
$\rm[Y\,II/Fe]$ & $-0.107 \pm 0.076$ & 3 \\
$\rm[Ba\,II/Fe]$ & $0.084 \pm 0.028$ & 1 \\
\enddata
\end{deluxetable*}

We show the position of \textit{Gaia} DR3 6769569470180424704 on an extinction-corrected color-magnitude diagram in the top panel of Figure~\ref{fig:galah_comparison}. As before, the extinctions are calculated based on reddenings retrieved from the 3D dust map of \citet{green_2019}. We confirm that the target is bluer than the main sequence of a 100 pc random comparison sample, but has typical colors for a metal-poor halo star. 

To check for any signatures of a double-lined spectroscopic binary, we follow the approach of \citet{nagarajan_symbiotic_2024} by comparing the high-resolution MIKE spectrum to objects with similar atmospheric parameters observed by the GALAH survey \citep{GALAH_2021}. Motivated by the derived stellar parameters in Table~\ref{tab:interesting}, we retrieve all GALAH spectra with $T_{\text{eff}} \in (5000, 7000)$ K, $\log{\left(g / \text{cm s$^{-2}$}\right)} \in (3.5, 4.5)$, and $[\text{Fe}/\text{H}] < -2.0$. We then degrade the MIKE spectrum to $R = 25,000$ and shift all spectra into their respective rest frames. We search for the closest matching GALAH spectra in pixel space by minimizing the total least-squares deviation in the wavelength range of $(6517\,\text{\AA}, 6690\,\text{\AA})$. We choose these bounds because virtually all of the signal in the overlapping region between the two spectra arises from the H$\alpha$ absorption line. We show the MIKE spectrum along with the closest matching GALAH spectrum in the bottom panel of Figure \ref{fig:galah_comparison}. We do not observe any strong emission lines or other unusual spectral features. The close match rules out the presence of a secondary that makes an appreciable contribution to the total light and has strong H$\alpha$ absorption. 

Based on the top ten matches in GALAH DR3, we empirically derive stellar parameters of $T_{\text{eff}} = 6120 \pm 130$ K, $\log{\left(g / \text{cm s$^{-2}$}\right)} = 4.2 \pm 0.2$, $\rm [Fe/H] = -2.30 \pm 0.23$, and $v_{\text{micro}} = 1.32 \pm 0.07$ km s$^{-1}$ for \textit{Gaia} DR3 6769569470180424704. While hot and metal-poor stars are relatively rare in GALAH DR3, these values are in $\sim2\sigma$ agreement with the best-fit derived parameters listed in Table~\ref{tab:mike_params}.

To summarize, if the \textit{Gaia} DR3 RV is correct, then \textit{Gaia} DR3 6769569470180424704 is very likely to host a dormant BH. However, we do not know if the DR3 RV is accurate, and further monitoring of the acceleration or orbital motion of this target is necessary to determine if this is the case. We check whether plausible orbital solutions that are consistent with the \textit{Gaia} DR3 RV, \textit{Gaia} DR3 RUWE, and epoch RVs exist, following the approach of M\"uller-Horn et al.\ (forthcoming). We find that only solutions with large companion masses, long orbital periods, and high eccentricities are consistent with the observed acceleration, though we defer further analysis of this to future work.

\section{Discussion} 
\label{sec:discussion}

\subsection{What are the false positives?}

False positives in our sample could be targets with spurious \textit{Gaia} DR3 mean RV measurements, or luminous binaries with primary-to-secondary flux ratios high enough to hide the signatures of the faint secondaries. However, we have removed clear double-lined spectroscopic binaries, hot stars, and fast rotators from consideration. While \textit{Gaia}'s RV processing pipeline could still potentially fail to recover the true mean RV for the targets in our vetted sample, it is clear from Figure~\ref{fig:compare_gaia_rvs} that the DR3 RVs are reliable for the vast majority of targets.

Luminous hierarchical triples are another possible source of false positives \citep[e.g.,][]{bashi_compact_2024}. Several objects in Table~\ref{tab:interesting} display RV variability that suggests a short orbital period, with a typical example being the candidate in the left panel of Figure~\ref{fig:multi_rv_figure}. While the large \texttt{RUWE} values for these targets are due to the motion of the outer tertiary, the large RV variability observed on short timescales is due to the orbital motion of the inner binary instead. Indeed, triple systems appear as contaminants whenever a sample of binaries is selected based on both high \texttt{RUWE} and large RV variability \citep[e.g.,][]{andrew_binary_2022}.

\subsection{How many low-metallicity stars in our sample have BH companions?}
\label{sec:hidden}

Metal-poor stars in the solar neighborhood are $\gtrsim 12$ Gyr old \citep[e.g.,][]{el-badry_where_2018, sesito_exploring_2021}. Using MIST isochrones \citep{choi_2016}, we find that metal-poor giants of this age cover a narrow mass range of $M = (0.78$--$0.80)\,M_{\odot}$. Assuming a Kroupa IMF \citep{kroupa_2001}, there are $\approx 172$ times more dwarfs with $M = (0.1$--$0.78)\,M_{\odot}$ compared to these giants. Assuming that Gaia BH3 is the only giant-BH system in our observed sample, and that BH companion probability is independent of luminous star mass in our mass range of interest, how many low-metallicity dwarfs in our sample are likely to have BH companions?

\begin{figure*}
    \centering
    \includegraphics[width=\textwidth]{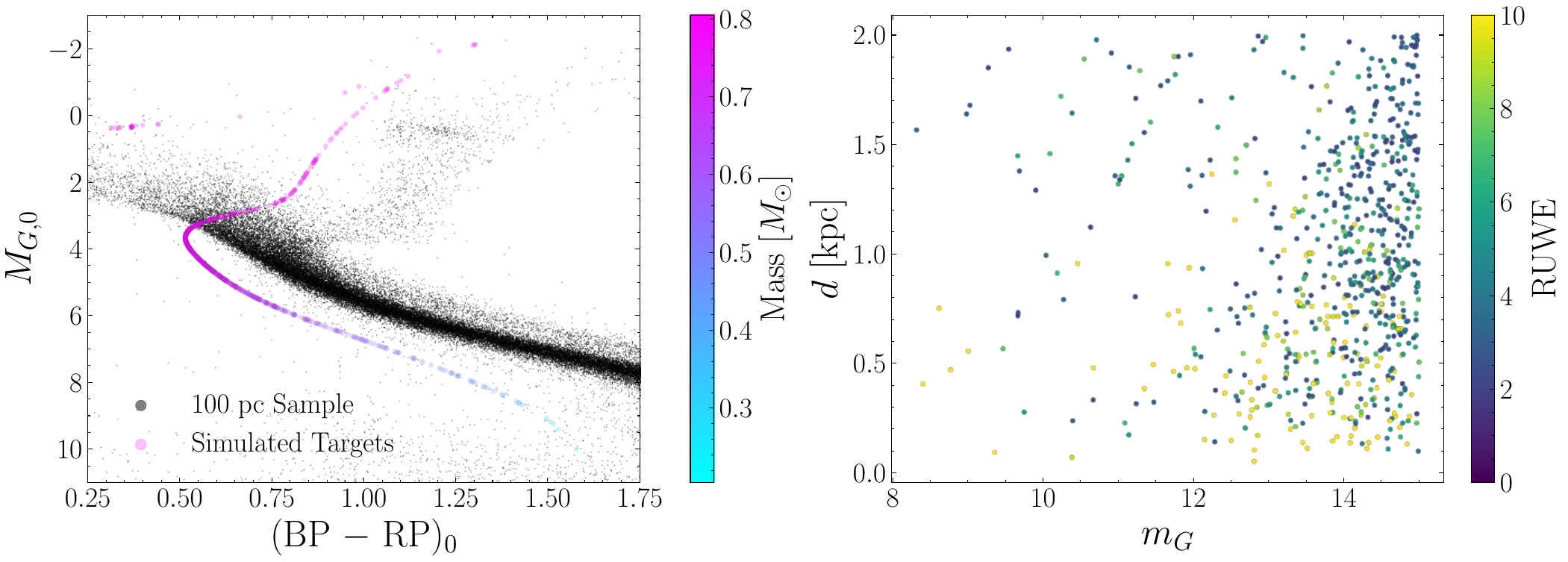}
    \caption{Properties of the simulated sample of low-metallicity BH binaries (Section~\ref{sec:hidden}). Left: Color-magnitude diagram, with a 100 pc random comparison sample shown for reference. The sources are colored by initial mass, and include both dwarfs and giants sampled from a 12 Gyr MIST isochrone with metallicity $\rm [Fe/H] = -2.0$. Right: Distance versus apparent magnitude for all simulated targets. The sources are colored by \texttt{RUWE}, with brighter and closer binaries displaying larger excess astrometric noise. We find that $\approx 8$ dwarfs make it into our sample for every giant.}
    \label{fig:sim_cmd}
\end{figure*}

To estimate the answer, we simulate a population of BH + luminous star binaries in the Galactic halo. We draw stars from a Kroupa initial mass function (IMF) \citep{kroupa_2001} between $0.1$ and $100\,M_{\odot}$. We sample Galactocentric radii from a power-law distribution, assuming a halo density profile $\rho(r) \propto r^{-3.5}$ (e.g., \citealt{xue_radial_2015, slater_2016, deason_apocenter_2018}; studies that adopt a double power-law profile find a shallower inner slope and steeper outer slope). We then calculate heliocentric distances and retain the stars that lie $< 2$ kpc away from the Sun, which we place at a distance of 8.122 kpc away from the Galactic Center \citep{gravity_2018} and at a height of 20.8 pc above the Galactic mid-plane \citep{bennett_jo_2019}. We estimate absolute magnitudes using synthetic photometry from a 12 Gyr MIST isochrone with initial metallicity $\rm [Fe/H] = -2.0$, and use them to calculate apparent $G$-band magnitudes. We assume zero extinction. We draw orbital periods from a log-uniform distribution between $10$ and $10^4$ days (reasonable for Gaia BH3-like systems formed in clusters, see e.g., \citealt{marin_pina_dynamical_2024}) and eccentricities from a thermal distribution. We assign sky locations by randomly sampling right ascensions and declinations from our catalog of candidate metal-poor binaries. Then, assuming that each luminous star has a $10\,M_{\odot}$ BH companion, we randomly sample binary orientations and calculate DR3 \texttt{RUWE} values using the \texttt{gaiamock} code \citep{el-badry_generative_2024}. Finally, we apply selection cuts of \texttt{RUWE} $> 2.0$ and $G < 15$ to identify the simulated sources that would make it into our sample. We display the properties of our simulated sample of targets in Figure~\ref{fig:sim_cmd}, confirming that brighter and closer binaries tend to have larger predicted \texttt{RUWE} values in DR3. 

Based on the location of the Hertzsprung Gap in the color-magnitude diagram in the left panel of Figure \ref{fig:sim_cmd}, we now define giants as stars with absolute $G$-band magnitude $M_{G, 0} < 2.5$ and dwarfs as stars with $M_{G, 0} > 2.5$. We find that there are $\approx 8$ times more dwarf-BH systems than giant-BH systems in the simulated low-metallicity sample; while the IMF predicts that there are more than 8 dwarfs for every giant, most of those dwarfs are too faint to make it into the sample. Thus, if Gaia BH3 is the only giant-BH system in our observed sample, we expect there to be $\approx 8$ undiscovered dwarf-BH systems in our observed sample as well. As discussed in Section~\ref{sec:sensitivity} below, we expect our search strategy to recover about half of these compact object binaries. If there are 5 genuine BH binaries in our sample, then the estimated false positive rate of our candidate sample is $62/67$, or about $93\%$.

These predictions depend upon Gaia BH3 being at a ``typical'' distance ($d = 0.6$ kpc) for low-metallicity BH binaries, and they could be overestimates if Gaia BH3 is a statistical anomaly. This history of astronomy has witnessed many cases where event rates and space densities based on the first-discovered object were overestimated \citep[e.g.,][]{schmidt_1963, abbott_2017}, so additional low-metallicity BH detections are needed to refine our predictions. 

\subsection{Search sensitivity}
\label{sec:sensitivity}

To characterize the efficacy of our search, we simulate the fraction of hidden BH companions recovered by our search strategy as a function of orbital period (log-spaced between $10^2$ and $10^4$ days) and BH mass (chosen from $5, 10, 20,$ or $50\,M_{\odot}$). We adopt the simulated Galactic population of low-metallicity binaries from Section \ref{sec:hidden}. For each binary brighter than $G < 15$, we assign sky locations by randomly sampling right ascensions and declinations from our catalog of candidate metal-poor binaries. Then, we randomly sample an eccentricity from a thermal distribution, assign a random binary orientation and phase, and mock observe the binary at a random time.

Using synthetic photometry from the MIST isochrone and Equations 2 and 3 in \cite{brown_dr2_2018}, we estimate the apparent $G_{\text{RVS}}$ magnitude of each binary. Then, we simulate a DR3 mean RV for each binary based on the \texttt{rv\_method\_used} parameter in DR3. In detail, for bright stars ($G_{\text{RVS}} \leq 12$), we compute the median of $20$ epoch RVs measured over a $10^3$ d baseline. This is motivated by the fact that the median \texttt{rv\_nb\_transits} in our observed sample is $\approx20$. For fainter stars ($G_{\text{RVS}} > 12$), we instead compute the simulated DR3 RV as the peak of a total cross-correlation function (CCF) constructed from combining 20 individual epoch CCFs. Here, each individual CCF is represented as a Gaussian with a width of $10$ km s$^{-1}$. This latter approach approximates the process actually carried out in DR3, implying that the RV reported in DR3 for faint stars is closer to the mode than to the mean of the epoch RVs.

\begin{figure*}
    \centering
    \includegraphics[width=\textwidth]{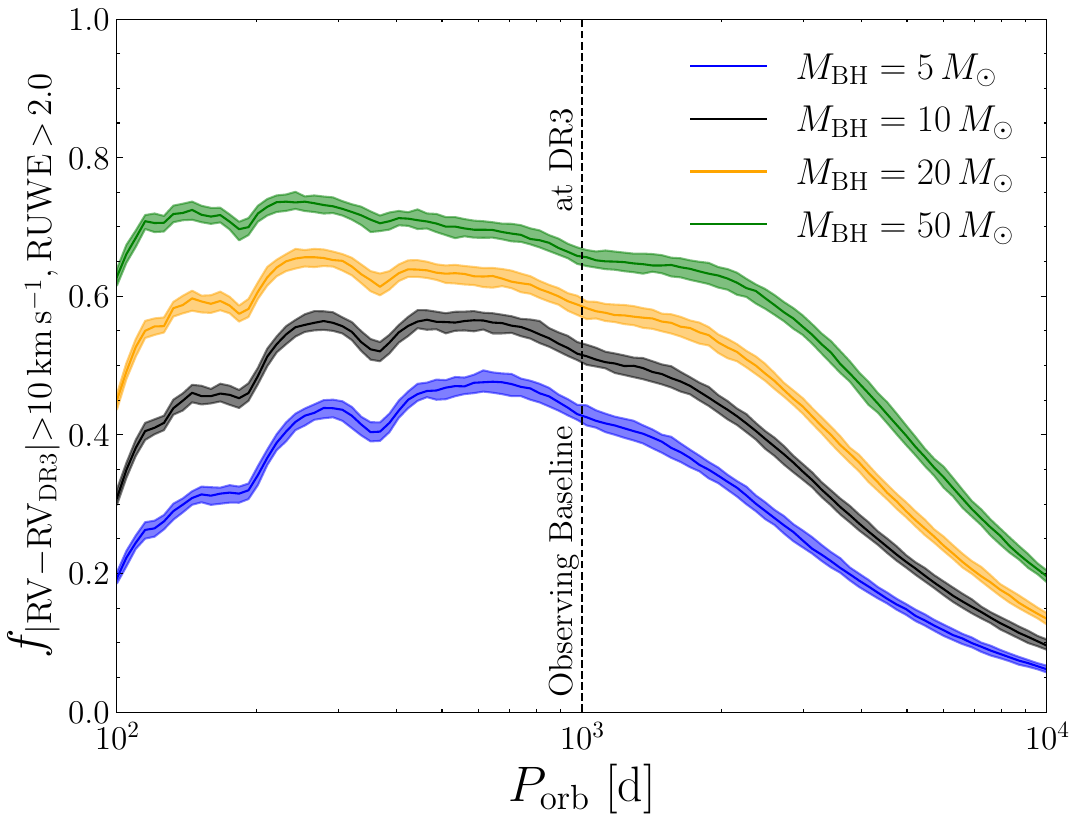}
    \caption{Median fraction of BH companions at $d < 2$ kpc and $G < 15$ (see Section~\ref{sec:hidden}) recovered by our search strategy as a function of orbital period and BH mass, with the shaded regions representing a $68\%$ confidence interval. For each binary, we sample eccentricities from a thermal distribution, assign a random orientation and phase, and mock observe the binary at a random time. We simulate their reported mean RVs by approximating the process carried out in DR3 (see Section \ref{sec:sensitivity}). At typical orbital periods and BH masses, we expect to recover at least half of the BHs as candidates, given detection thresholds of an RV discrepancy $> 10$ km s$^{-1}$ and \texttt{RUWE} $> 2.0$. Some real BH binaries may not be recovered if they have a face-on orientation or are observed at a time during which their epoch RV is similar to their average RV. We find that the search sensitivity peaks at orbital periods of a few hundred days, and is significantly lower at much shorter or longer orbital periods.}
    \label{fig:recover_sim}
\end{figure*}

Using \texttt{gaiamock}, we compute the fraction of mock observed BH binaries that have both predicted DR3 \texttt{RUWE} $> 2.0$ and a discrepancy between their epoch RV and mean DR3 RV greater than our adopted threshold of $10$ km s$^{-1}$. To derive the uncertainty on this fraction, we repeat our simulation $100$ times. We plot our results in Figure~\ref{fig:recover_sim}. We find that, assuming typical orbital periods ($\sim 10^3$ d) and BH masses ($\sim 10\,M_{\odot})$, our search strategy recovers about half of BHs in low-metallicity binaries that lie closer than $2$ kpc and have luminous companions brighter than $G = 15$. Some real BH binaries may not be recovered if they have orientations close to face-on or are observed at a time during which their epoch RV is similar to their average RV. The search sensitivity peaks at orbital periods of a few hundred days, with the sensitivity being significantly lower at much shorter or longer orbital periods.

\section{Conclusion}
\label{sec:conclusion}

The discovery of Gaia BH3 in DR4 pre-release astrometry suggests that luminous companions to Gaia BHs may be dramatically overrepresented at low metallicities. To investigate this hypothesis, we have initiated a radial velocity (RV) survey of metal-poor stars with significant astrometric excess noise (\texttt{RUWE}) values in \textit{Gaia} DR3 using the FEROS and APF spectrographs. We compared these epoch RVs against the mean RV values reported for these targets in DR3 to find potential black hole (BH) and neutron star (NS) candidates. The first results of our survey are reported here; continued monitoring of promising candidates is ongoing. We summarize our main conclusions below.

\begin{itemize}
    \item \textit{Experimental design:} In Gaia BH3, the red giant's RV is different from the mean DR3 RV by at least 20 km s$^{-1}$ over $\approx70\%$ of its orbit, and by at least 10 km s$^{-1}$ over $\approx85\%$ of its orbit, suggesting that a search strategy for BHs and NSs based on this discrepancy would be effective (Figure~\ref{fig:bh3_rv_astro}). Furthermore, for simulated low-metallicity dwarf-BH and giant-BH binaries, the predicted \texttt{RUWE} in DR3 is significantly above a detectability threshold of $2.0$ for typical periods of $100$--$5000$ d and distances $< 2$ kpc (Figure~\ref{fig:ruwe_trend_fig}). Hence, to search for dormant BHs, we measure instantaneous RVs for low-metallicity stars with elevated \texttt{RUWE} values in DR3 and check whether they are highly discrepant with their corresponding mean DR3 RVs.
    \item \textit{Summary of the sample:} We select a sample of bright, metal-poor sources with $\rm [M/H] < -1.0$ and excess astrometric noise in DR3, adopting the robust, data-driven metallicities derived from XP spectra by \citet{andrae_2023}. We use two different selection cuts (Section~\ref{sec:sample}) for targets with $\rm [M/H] < -1.5$ and \texttt{RUWE} $> 2.0$ and targets with $\rm [M/H] < -1.0$, \texttt{RUWE} $> 1.4$, and large estimated $a_0$, respectively. We apply additional quality cuts to exclude marginally resolved binaries and stars with disk-like orbits. We obtained at least one FEROS or APF spectrum for 528 targets out of a sample of 717 candidates, with the remainder scheduled to be observed in the future. The sample of observed targets has $\rm[M/H] < -1.0$, \texttt{RUWE} $> 1.4$, $G = 9$--$15$, and is bluer than the main sequence on a color-magnitude diagram (Figure~\ref{fig:sample_properties}). 
    \item \textit{RV measurements:} We measure epoch RVs by minimizing the chi-squared statistic between the target spectrum and the BOSZ template spectrum with stellar parameters closest to those derived from XP spectra for the target by \citet{andrae_2023} (Figure~\ref{fig:good_chi_squared}). Some spectra have low SNR or are contaminated by moonlight, and others show either a lack of prominent spectral lines or strong evidence of two luminous components (Figure~\ref{fig:bad_chi_squared}). We flag these spectra and remove the corresponding RVs from further consideration. All of our RVs are listed in Table~\ref{tab:comparison}.
    \item \textit{BH/NS candidates:} We compare all of our derived RVs against the mean RVs reported in DR3, finding that most are in good agreement (Figure~\ref{fig:compare_gaia_rvs}). In several cases, the RVs are discrepant by more than 10 km s$^{-1}$, indicating the potential for a massive, dark companion. We list all promising candidates in Table~\ref{tab:interesting}. We obtained additional follow-up spectra for promising targets, but have not yet fully solved the orbits of any candidates (Figure~\ref{fig:multi_rv_figure}). Some targets show short-timescale RV variability, and are likely to be hierarchical triples. Extended RV monitoring will be vital to determine the true nature of these BH or NS candidates.
    \item \textit{A promising BH candidate:} We identify \textit{Gaia} DR3 676956947018042470 as having the largest discrepancy ($\approx 98$ km s$^{-1}$) between its epoch RVs and its mean RV reported in DR3. We obtained several follow-up spectra, including a high-resolution MIKE spectrum, to investigate this promising BH candidate (Figure~\ref{fig:interesting_object}). From the MIKE spectrum, we infer best-fit stellar parameters and abundances for the target, finding that they are typical for a low-metallicity halo star given our selection criteria (Figure~\ref{fig:galah_comparison}). We measure a slow acceleration of $(-11 \pm 2)$ km s$^{-1}$ yr$^{-1}$ away from the mean DR3 RV, but without knowledge of whether the reported mean RV is accurate, further monitoring is necessary to determine the candidate's true nature. 
    \item \textit{Simulations:} To characterize the effectiveness of our search strategy, we mock-observe a simulated population of BH + luminous star binaries in the Galactic halo, finding that if Gaia BH3 is the only BH + giant binary in our sample, then we expect there to be $\approx 8$ undiscovered BH + dwarf binaries in our sample as well (Figure~\ref{fig:sim_cmd}). We also simulated the fraction of BH companions recovered as a function of orbital period and BH mass, finding that we expect to detect at least half of BH companions with periods of order $10^3$ d (Figure~\ref{fig:recover_sim}).
\end{itemize}

We will continue to follow-up the interesting objects identified in Table~\ref{tab:interesting}. Since these targets have acceleration solutions or high \texttt{RUWE} values rather than orbital solutions in DR3, the orbital periods of these systems are likely to be on the order of $1$--$10$ years \citep{el-badry_generative_2024}. This renders it practically challenging to cover the full RV curves of these targets with spectroscopic follow-up. 

However, we will obtain another epoch RV for these targets following the release of DR4, in which many of these objects are likely to receive an orbital solution. This will provide us with at least two RV measurements separated by a multi-year time baseline. We will fit these RVs together with the epoch astrometry in DR4 to constrain the parameters of the low-metallicity binaries with high implied mass functions and uncover any hidden NSs or BHs.

\begin{acknowledgments}
This research was supported by NSF grant AST-2307232. HWR acknowledges the European Research Council for support from the ERC Advanced Grant ERC-2021-ADG-101054731. CYL acknowledges support from the Harrison and Carnegie Fellowships. This work has made use of data from the European Space Agency (ESA) mission
{\it Gaia} (\url{https://www.cosmos.esa.int/gaia}), processed by the {\it Gaia}
Data Processing and Analysis Consortium (DPAC,
\url{https://www.cosmos.esa.int/web/gaia/dpac/consortium}). Funding for the DPAC
has been provided by national institutions, in particular the institutions
participating in the {\it Gaia} Multilateral Agreement. This paper includes data gathered with the 6.5 meter Magellan Telescopes located at Las Campanas Observatory, Chile. 
\end{acknowledgments}

\begin{contribution}

PN was responsible for performing the data analysis. KE came up with the initial research concept and obtained the funding. HR measured abundances from the high-resolution spectra. CYL, JDS, HI, and JRL assisted in planning and executing the APF observations, while JMH, RS, HWR, and VC assisted in planning the FEROS observations. JDS obtained and reduced the MIKE spectrum. RA measured metallicities from XP spectra. All authors were involved in writing and editing the manuscript.


\end{contribution}

%
\facilities{APF, Max Planck:2.2m (FEROS), Magellan:Baade (MIKE)}

\software{astropy \citep{2013A&A...558A..33A, 2018AJ....156..123A, 2022ApJ...935..167A}, \texttt{emcee} \citep{Foreman-Mackey_2013}}


\clearpage

\appendix

\section{Radial Velocities}
\label{sec:all_rvs}

We provide a summary of all of our measured radial velocities (RVs) in Table~\ref{tab:comparison}. We group RVs by target and sort by largest RV discrepancy with the mean RV reported in DR3. We flag unreliable RVs derived from spectra that have either low signal-to-noise (SNR) values $< 2$ at $5200\,\text{\AA}$ or moonlight contamination that causes the barycenter correction (BC) to be within $10$ km s$^{-1}$ of the derived RV. We also flag RVs derived from spectra visually identified to be of hot stars (hot), fast rotators ($v \sin i$), or spectroscopic binaries (SB2).

\startlongtable
\begin{deluxetable*}{ccccccc}
\tablecaption{Derived radial velocities (RVs) for all observed targets. We also provide the mean RV reported in \textit{Gaia} DR3 for comparison. \label{tab:comparison}}
\tablehead{\colhead{\textit{Gaia} DR3 ID} & \colhead{JD} & \colhead{RV} & \colhead{Instrument} & \colhead{\textit{Gaia} DR3 Mean RV} & \colhead{Flags} & \colhead{RV Shift} \\
& d & km s$^{-1}$ & & km s$^{-1}$ & & km s$^{-1}$ \\
\colhead{(1)} & \colhead{(2)} & \colhead{(3)} & \colhead{(4)} & \colhead{(5)} & \colhead{(6)} & \colhead{(7)}}
\startdata
31530905403904512 & 2460559.9042 & $-59.67 \pm 0.11$ & FEROS & $-58.95 \pm 1.94$ &  & $0.72 \pm 1.95$ \\
31530905403904512 & 2460570.8191 & $-65.16 \pm 0.14$ & FEROS & $-58.95 \pm 1.94$ &  & $6.21 \pm 1.95$ \\
75519548132523136 & 2460569.8113 & $58.07 \pm 0.20$ & FEROS & $50.99 \pm 4.27$ &  & $7.08 \pm 4.28$ \\
187869776566408960 & 2460575.8128 & $25.09 \pm 0.24$ & APF & $27.81 \pm 6.21$ &  & $2.72 \pm 6.21$ \\
207547530009853952 & 2460574.8033 & $11.79 \pm 0.26$ & APF & $13.02 \pm 3.99$ &  & $1.23 \pm 4.00$ \\
249249188555702784 & 2460563.9349 & $-285.39 \pm 0.19$ & APF & $-276.70 \pm 4.36$ &  & $8.69 \pm 4.36$ \\
290511523477577600 & 2460553.9909 & $-48.02 \pm 0.17$ & APF & $-41.60 \pm 3.33$ &  & $6.42 \pm 3.34$ \\
316060462855379840 & 2460566.8837 & $-216.74 \pm 0.20$ & APF & $-214.23 \pm 3.83$ &  & $2.51 \pm 3.84$ \\
\enddata
\end{deluxetable*}


\bibliography{bibliography}{}
\bibliographystyle{aasjournalv7_modified}



\end{document}